\documentclass[iop]{emulateapj}

\newcommand{\expnt}[2]{\ensuremath{#1 \times 10^{#2}}}   
\newcommand{\rxjw}{RX~J1856.5$-$3754}
\newcommand{\rxjk}{RX~J0720.4$-$3125}
\newcommand{\RXJ}{RX~J0806.4$-$4123}
\newcommand{\rbs}{RX~J1308.6+2127}
\newcommand{\rbsb}{RX~J2143.0+0654}
\newcommand{\rxjvk}{RX~J1605.3+3249}
\newcommand{\rxjb}{RX~J0420.0$-$5022}
\newcommand{\cxo}{{\em Chandra}}

\newcommand{\xmm}{{\em XMM}}
\newcommand{\hst}{{\em HST}}
\defcitealias{kkvk02}{KKvK02}
\defcitealias{kkvk03}{KKvK03}
\defcitealias{kvkm+03}{K+03}
\defcitealias{vkk01}{vKK01}

\shorttitle{Counterparts and SEDs of Nearby Neutron Stars}
\shortauthors{Kaplan et al.}

\begin{document}
\title{New Optical/UV  Counterparts and the Spectral Energy
 Distributions of Nearby, Thermally Emitting, Isolated Neutron Stars}

\author{D.~L.~Kaplan\altaffilmark{1}, A.~Kamble\altaffilmark{1}, M.~H.~van
  Kerkwijk\altaffilmark{2}, and W.~C.~G.~Ho\altaffilmark{3}}
\altaffiltext{1}{Physics Dept., U. of Wisconsin - Milwaukee, Milwaukee
  WI 53211; kaplan@uwm.edu, kamble@uwm.edu}

\altaffiltext{2}{Department of Astronomy and Astrophysics, University
  of Toronto, 50 St.\ George Street, Toronto, ON M5S 3H4, Canada;
  mhvk@astro.utoronto.ca}

\altaffiltext{3}{School of Mathematics, University of Southampton,
  Southampton, SO17 1BJ, United Kingdom; wynnho@slac.stanford.edu}

\slugcomment{Draft \today}

\begin{abstract}
 We present \textit{Hubble Space Telescope} optical and ultraviolet
 photometry for five nearby, thermally emitting neutron stars.  With
 these measurements, all seven such objects have confirmed optical and
 ultraviolet counterparts.  Combining our data with archival
 space-based photometry, we present spectral energy distributions for
 all sources and measure the ``optical excess'': the factor by which
 the measured photometry exceeds that extrapolated from X-ray spectra.
 We find that the majority have optical and ultraviolet fluxes that
 are inconsistent with that expected from thermal (Rayleigh-Jeans)
 emission, exhibiting more flux at longer wavelengths.  We also find
 that most objects have optical excesses between 5 and 12, but that
 one object (\rbsb) exceeds the X-ray extrapolation by a factor of
 more than 50 at 5000\,\AA, and that this is robust to uncertainties
 in the X-ray spectra and absorption.  We consider explanations for
 this ranging from atmospheric effects, magnetospheric emission, and
 resonant scattering, but find that none is satisfactory.
\end{abstract}


\section{Introduction}

Among the nearby neutron stars known, seven show emission that appears
predominantly thermal,\footnote{Fainter candidate members of the same
  class have been identified by, e.g., \citet{pmt+09}.} with inferred
temperatures of $\sim\!10^6\,$K.  These so-called isolated neutron
stars (INSs) are young, $\lesssim\!1\,$Myr old, and the thermal
emission is thought to be due to residual heat as their X-ray
luminosities are considerably more than their spin-down luminosities
$\dot E$.  They differ from similarly aged pulsars not only in the
absence of non-thermal radio and X-ray emission, but also in their
long, 3--10\,s spin periods and large, $\sim10^{13}\,$G magnetic
fields (for reviews, \citealt{haberl07,vkk07,kaplan08}).

The INSs have attracted much attention, in part because of the hope
that their properties could be used to constrain the poorly understood
behavior of matter in their ultra-dense interiors: since the details
of the neutron star's interior affect its radius \citep{lp07} and
cooling history \citep{yp04}, the wide range of theoretical
possibilities can be limited with data.  In this respect, the thermal
emission from INSs is particularly interesting, since one might derive
constraints on mass, radius, and cooling history from spectral
properties such as effective temperature, surface gravity, and
gravitational redshift.

Progress has been stymied, however, by difficulties in interpreting
the observed spectra: at present, both the composition and state of
matter in the photosphere remain unknown, with hydrogen, helium, and
mid-Z elements (O, N, Ne, etc.) in states ranging from gaseous to
condensed all being considered (see, e.g., contributions to
\citealt*{ptz06}).  Below, we use the brightest and best studied INS
to illustrate those problems and the part played by optical and
ultraviolet measurements.

\subsection{The Puzzle of \rxjw}
The emission of INSs was known to be roughly thermal, but it came as a
major surprise when long {\em Chandra} and {\em XMM} observations
found that the spectrum of the brightest INS, \rxjw, could be
reproduced with a featureless black body
\citep[e.g.,][]{bzn+01,bhn+03}.  The reason for the surprise was that,
for a light element (H or He) atmosphere, the spectrum may be
featureless but the high-energy tail should be harder than a Wien tail
(because opacity decreases towards higher energies and hence one
should see deeper, hotter layers), while for heavier elements, the
overall shape may be blackbody-like but spectral features should be
present.

\setlength{\tabcolsep}{2pt}
\begin{deluxetable}{c c c c c}
\tablewidth{0pt}
\tablecaption{Summary of \hst\ optical and UV observations\label{tab:HST_obs}}
\tablehead{
\colhead{RX~J} & \colhead{Instrument and Filter} & \colhead{Date} &
\colhead{Exp.} & \colhead{ST Mag.\tablenotemark{a}}\\
& & \colhead{(UT)} & \colhead{(s)} 
}
\startdata
0420.0$-$5022	&	ACS/WFC F475W	& 2009-10-24	&	2424	& $27.85\pm0.25$	\\
				&	ACS/SBC 	 F140LP	& 2009-12-16	&	2856	& $22.85\pm0.08$	\\
0806.4$-$4123 	&	ACS/WFC F475W	& 2010-05-19	&	4868	& $27.92\pm0.22$	\\
				&	ACS/SBC 	 F140LP	& 2009-12-15	&	5692	& $23.61\pm0.11$	\\
1308.6+2127	&	ACS/WFC F475W	& 2009-08-08	&	4676	& $27.97\pm0.25$	\\
				&	ACS/SBC 	 F140LP	& 2009-08-06	&	5504	& $23.79\pm0.09$	\\
1605.3+3249 	& ACS/WFC F606W & 2005-02-06 & 4728  & $27.26\pm0.07$\\
 	&ACS/WFC F475W	& 2009-08-01	&	2256	& $26.61\pm0.09$	\\
				&	ACS/SBC 	 F140LP	& 2009-01-18	&	2688	& $22.62\pm0.07$	\\	
2143.0+0654 	&	ACS/WFC F475W	& 2010-05-19	&	7076	& $26.31\pm0.05$	\\
				&	ACS/SBC 	 F140LP	& 2010-05-23 	&	8296	& $23.38\pm0.06$	\\
\enddata
\tablenotetext{a}{Values are magnitudes in the STMAG system that have
 been corrected for finite apertures. Aperture
 corrections amounted to $0.18$\,mag for the F475W/F606W data and
 0.33\,mag for the F140LP data.  We estimate an additional 5\%
 systematic uncertainty on the SBC/F140LP photometry owing to
 uncertainties in the aperture correction.  The zero points used were
 26.67, 25.75, and 20.316 for F606W, F475W, and F140LP, respectively,
 taken from the revised calibration for ACS given at
 \url{http://www.stsci.edu/hst/acs/analysis/zeropoints/}}
\end{deluxetable}
\setlength{\tabcolsep}{5pt}

When combined with the parallax determined from {\em HST} observations
(\citealt{wal01}; \citealt*{kvka02}; \citealt{wl02,wel+10}), the
implied blackbody radius was too small for a neutron star, leading to
speculation that the object might be a ``quark star'' \citep{dmd+02}.
The suggestion of a quark star, however, ignored that the optical and
ultraviolet measurements were inconsistent with a single blackbody,
being in excess by a factor $\sim\!8$ over the extrapolation of the
X-rays \citep[][and see Fig.~\ref{fig:ox}]{wm97,vkk01}.  This excess
has been taken as evidence for two regions on the surface: a hotter
one primarily responsible for the X-ray emission, and a larger, cooler
one responsible for the optical \citep[e.g.,][]{br02}.  In the context
of this model, the radius is not small but rather puzzlingly large.
Furthermore, the amplitude of the X-ray pulsations is surprisingly
tiny \citep{tm07}.

As an alternative, it has been suggested that the surface is
condensed, but covered by a thin gaseous layer of hydrogen (see
\citealt{hkc+07}; also \citealt*{mzh03,ztd04}).  Hydrogen can unify the
optical and X-ray emission, since compared to a blackbody, a hydrogen
atmosphere will appear to show an optical excess by an amount that
depends on the magnetic field strength \citep*{hpc08}, and the
pulsations will be determined by non-uniformities in the temperature
\citep{ho07}.  For a thick hydrogen atmosphere one expects a
high-energy tail, but a suitably thin atmosphere will be transparent
at high energies, allowing one to see the blackbody-like emission from
the condensed surface below.

By construction, the above model reproduces the spectrum.  But it has
other advantages: (i) It resolves the problem of the small radius:
because of its non-gray opacities, the temperature is smaller and the
radius larger than for a pure blackbody model; (ii) For the dipolar
magnetic field of $1.5\times10^{13}\,$G \citep{vkk08}, a heavy-element
surface could indeed be condensed \citep{ml07b}, though we note that
the field is stronger than inferred by \cite{hkc+07}; (iii) For other
INSs, hydrogen might be responsible for the observed spectral features
(see \citealt{haberl07,vkk07}); and (iv) The appearance of a thin
layer of hydrogen may be the easiest explanation for the change in
spectrum observed for \rxjk\ \citep{vkkpm07}.

While promising, this model arguably is contrived, and certainly for
\rxjw\ it leaves it unclear how to proceed to test it.  Fortunately,
however, the other INSs provide much more information.

\subsection{Expectations for the other Isolated Neutron Stars}

For the hydrogen models described above -- and generally for
atmospheric models -- the optical excess depends on the magnetic field
strength.  For \rxjw, the field could only be constrained by the
absence of X-ray absorption features, but for other sources the
situation is better: these have absorption features, which any model
will have to reproduce simultaneously with the optical excess, and at
field strengths consistent with those derived from timing.

For all INSs, optical counterparts have been searched for in deep
optical observations, leading to four secure identifications and two
likely identifications.  So far, apart from \rxjw, only for the
second-brightest INS, \rxjk, has the full optical to UV range been
probed.  The interpretation is difficult, though, partly because the
fluxes are somewhat inconsistent with those of a Rayleigh-Jeans spectrum
\citep{kvkm+03,mzh03}, and partly because, uniquely among the INSs, its
X-ray spectrum changed (which poses problems and possibilities all of
its own; \citealt{dvvmv04,htdv+06,vkkpm07,hhv+09}).

The possible presence of non-thermal emission in the optical, also
suggested for \rxjvk\ \citep{msh+05,zdlmt06}, makes it difficult to
interpret the optical emission.  However, for \rxjk\ and \rxjw\ the
ultraviolet emission does seem completely thermal.  This is similar to
what is found for comparably-aged radio pulsars such as PSR~B0656+14
(\citealt*{pwc97}; \citealt{ssl+05}) and Geminga \citep{kpzr05}.
(Note, though, that for those pulsars the ultraviolet emission is
generally consistent with the X-ray extrapolation.)

Given the above, we obtained \hst\ optical and UV photometry for those
five INSs that had not been studied in detail before.  We complemented
this with other \hst\ photometry for the INSs (reanalyzing the data
where necessary); we restrict our analysis to \hst\ data both for the
high signal-to-noise that it often implies as well as the uniform
quality of the calibration.

In Section~\ref{sec:obs} we present our new data, and give details
concerning the identification and photometry of the
counterparts.  In Section~\ref{sec:cpt}, we compare our results to
previous (usually ground-based) searches, and perform detailed
spectral fitting of the optical/UV data for all 7 INSs, with and
without reference to the X-ray spectra.  We present our discussion and
conclusions in Section~\ref{sec:disc}.

\section{Observations and Analysis}
\label{sec:obs}
Of the seven confirmed INSs discovered by \textit{ROSAT}, three do not
have confirmed optical counterparts: for \rxjb\ and \rbs, the
associations are based only on positional coincidence, without
confirmation based on spectrum or proper motion.  Furthermore, \rxjvk\
and \rbsb\ did not have ultraviolet photometry.  We observed all of
these with \hst\ in the optical and ultraviolet.  A log of the
observations and our photometry are given in Table~\ref{tab:HST_obs}.

\begin{figure*}
\plottwo{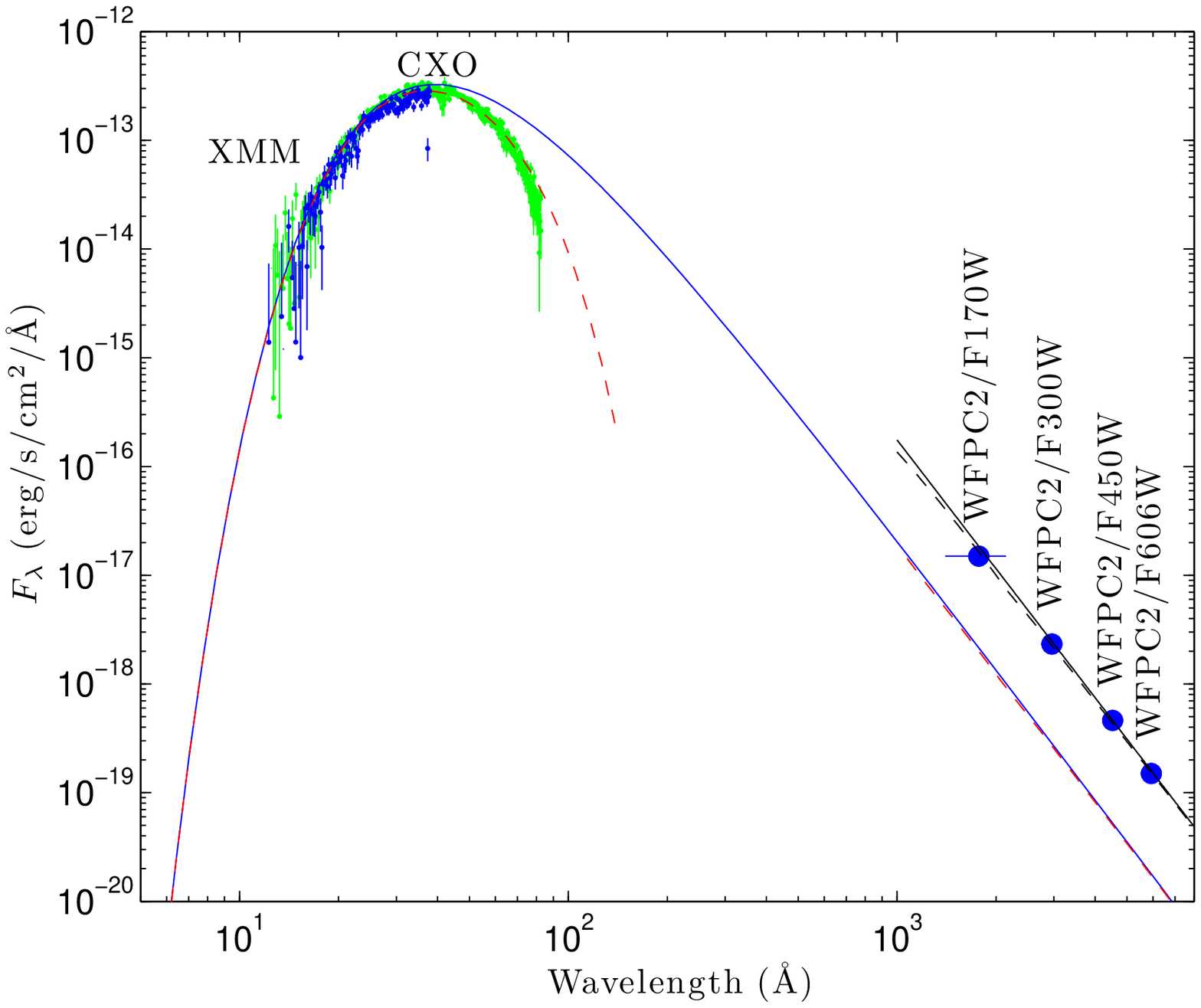}{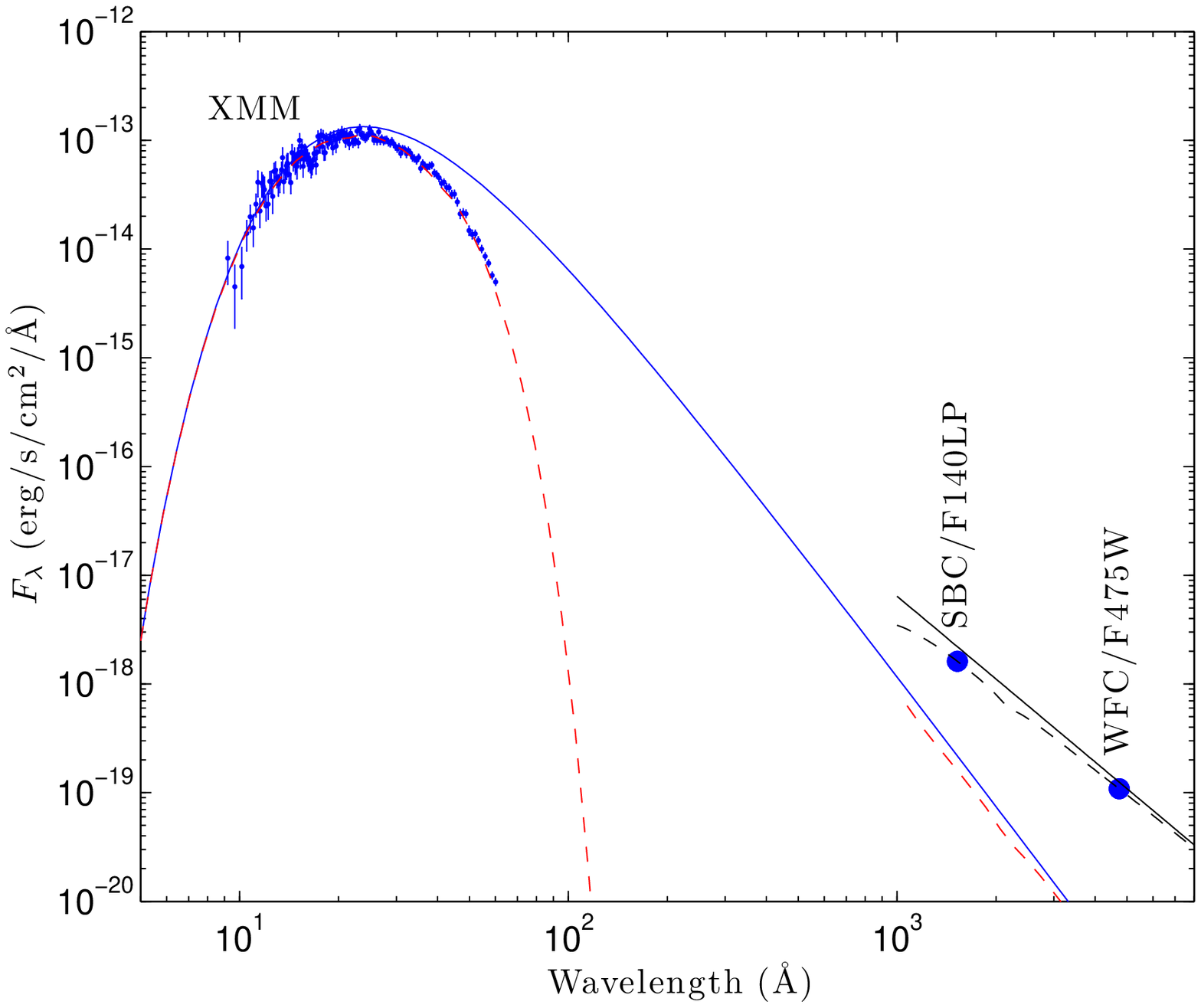}
\caption{X-ray to optical SEDs of \rxjw\ (left) and \rbsb\ (right). We
plot archival X-ray data from \textit{XMM} (blue) and \textit{CXO}
(green) along with the best-fit models (dashed red lines); the model
for \rxjw\ is just a blackbody \citep{bhn+03} while that for
\rbsb\ includes a broad absorption line \citep{kvk09}.  The unabsorbed
models are the solid blue lines.  In the optical/UV we plot the
\textit{HST} data from \citet{vkk01} and this paper along with the
best-fit power-law: the absorbed power-law is the black dashed line,
while the unabsorbed power-law is the solid black line.}
\label{fig:ox}
\end{figure*}

\subsection{Image Reduction and Combination}
The data for each of the five targets consist of multiple exposures
covering the SDSS $g^\prime$ band (F475W filter, centered near
4700\,\AA) with the Advanced Camera for Surveys (ACS) Wide Field
Channel (WFC) and the near-UV (F140LP filter, centered near 1500\,\AA)
with the ACS Solar Blind Channel (SBC).  For both the WFC and SBC data
we used four sub-pixel dithered exposures per orbit and 1--3 orbits
per object; for the WFC the object was placed at the center of the
WFC1 detector so that we did not have to deal with the gap between the
detectors.  For completeness, we also re-analyzed the ACS/WFC F606W
data on \rxjvk, since no photometry is given in the original
publication by \citet{zdlmt06}.

WFC images obtained after Servicing Mission 4 (SM4) show charge trails
along the CCD columns and faint stripes along the rows.  The charge
trails are due to the CCDs' degrading charge transfer efficiency
(CTE), while the stripes reflect a problem in the new electronics.
These artifacts can significantly affect the detection of very faint
sources as well as photometry and astrometry.  We used the tasks
\texttt{acs\_destripe} (version 0.2.1) and \texttt{CteCorr} (also
version 0.2.1, with the reference file \texttt{pctefile\_101109.fits};
see \citealt{ab10}) to correct the images, finding noticeably
improvements in appearance.  Checking the resulting photometry against
that of raw images, we confirmed the conclusions of \citet{ab10} that
their algorithm restores photometric accuracy.  Comparing against the
average CTE corrections computed by \citet{clkp+09}, we found good
agreement, although with some scatter (again, similar to what was
found by \citealt{ab10}).  For the F606W data on \rxjvk, taken before
SM4, the stripes are not present and we could not use the pixel-based
CTE correction since the data were taken before SM4 and would require
different calibration parameters, but other methods are possible to
correct the reduced CTE losses.  The SBC data do not suffer from such
artifacts.

For each source in Table~\ref{tab:HST_obs}, `destriped' and
CTE-corrected (as necessary) images were drizzled using
\texttt{multidrizzle} \citep{kfhh02} onto a single image. We
experimented with the drizzle parameters\footnote{Following
  \url{http://www.stsci.edu/hst/HST\_overview/documents/multidrizzle/multidrizzle.pdf}.}
to balance resolution and uniformity of sampling, settling on a set
that gave good image quality: for the WFC data we used a final pixel
scale of $0.04\arcsec {\rm pixel}^{-1}$ and a pixel fraction of 0.9,
while for SBC we used a pixel scale of $0.03\arcsec {\rm pixel}^{-1}$
and a pixel fraction of 0.9.  We verified that our photometry did not
depend on the choices that we made.

\subsection{Astrometry and Counterpart Identification}
For all INSs, the SBC images show an obvious, rather bright
counterpart, with at best a few other faint sources present.  In
contrast, in the WFC images, the counterparts are faint and there are
many other, brighter stars.  Previous identification of optical
counterparts to the INSs has relied on absolute astrometry using the
positions measured in X-rays and approximate ties between the X-ray
and optical coordinate systems \citep[e.g.,][]{kkvk03}.  This can lead
to uncertainties, as the absolute X-ray positions are typically
accurate to only $0\farcs6$ or so.  However, the SBC images allow us
to conclusively identify the counterparts with much higher
accuracy. To do so we started by registering the SBC and WFC images.
In two cases, we estimated offsets from single, bright optical stars
that had faint ultraviolet counterparts, while in the others we relied
on diffuse emission from galaxies (smoothing the SBC and WFC data as
necessary to isolate a single bright component).

In all cases the registration was unambiguous, and led to clear
identifications of the optical counterparts to the INSs.  The SBC and
WFC positions matched to within $<0\farcs04$ (since they were taken at
most 7\,months apart, a typical proper motion of $100\,{\rm
 mas\,yr}^{-1}$ would lead to a comparable intrinsic offset).

\begin{figure*}
\centerline{\includegraphics[width=0.25\textwidth]{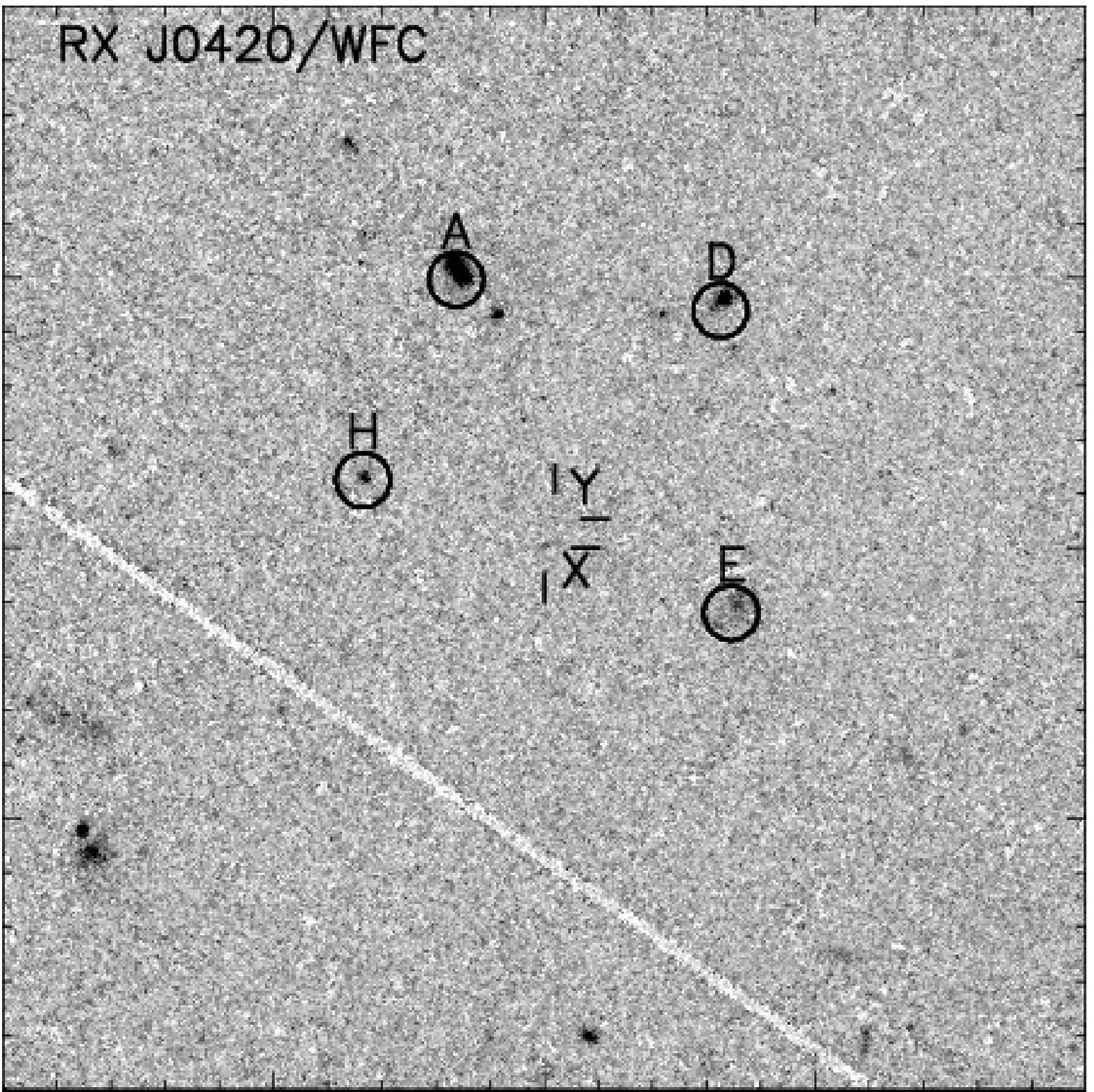}%
\includegraphics[width=0.25\textwidth]{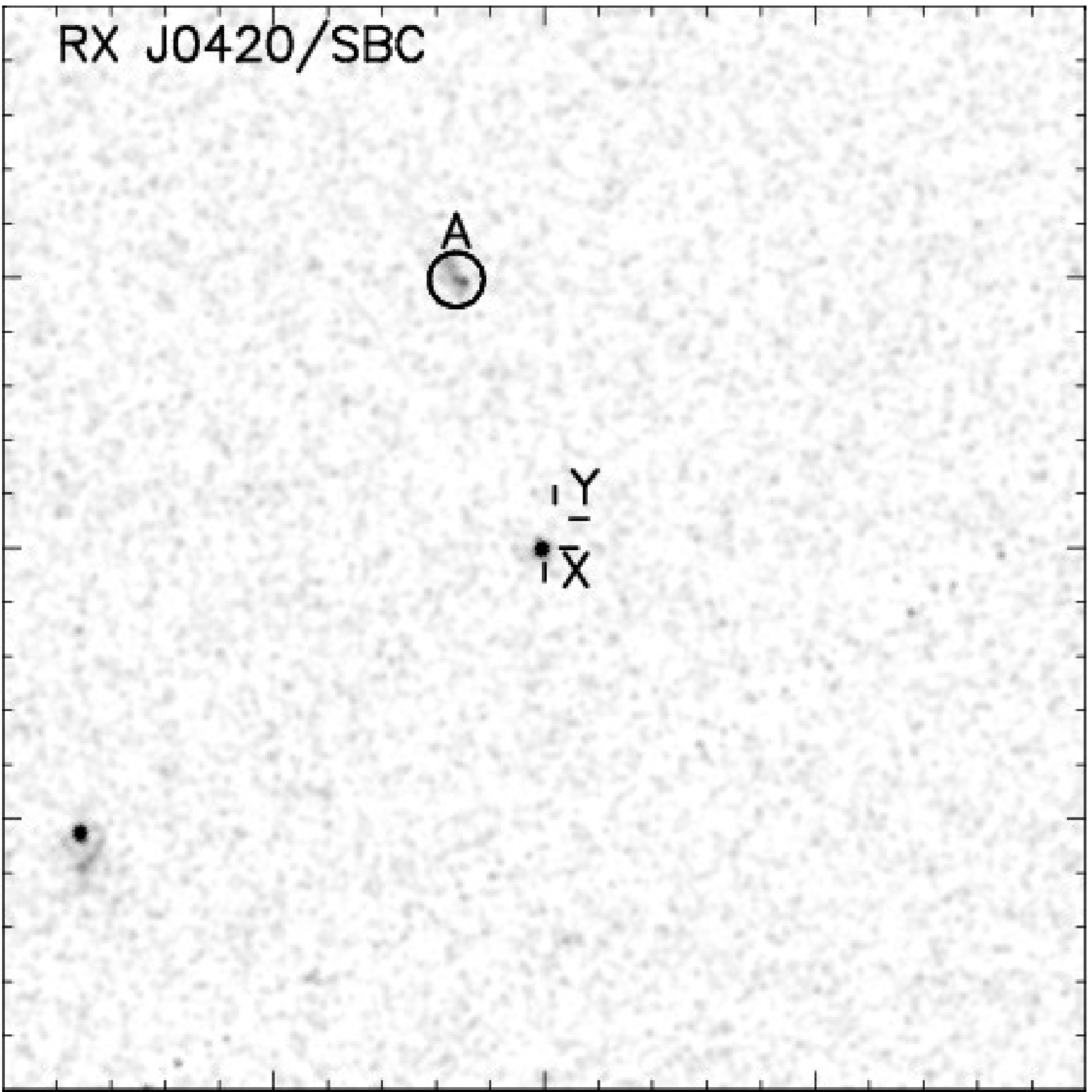}\includegraphics[width=0.25\textwidth]{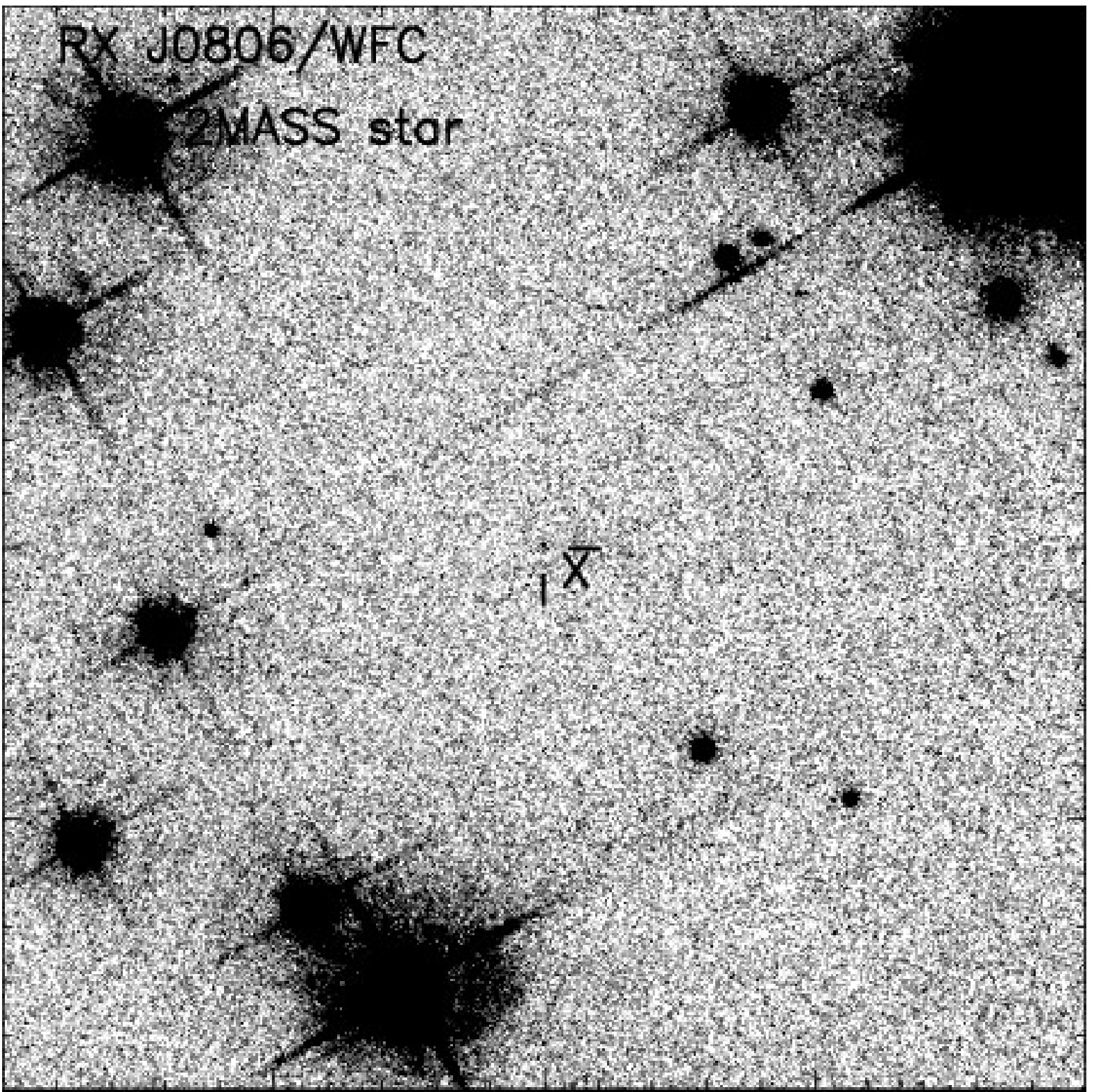}%
\includegraphics[width=0.25\textwidth]{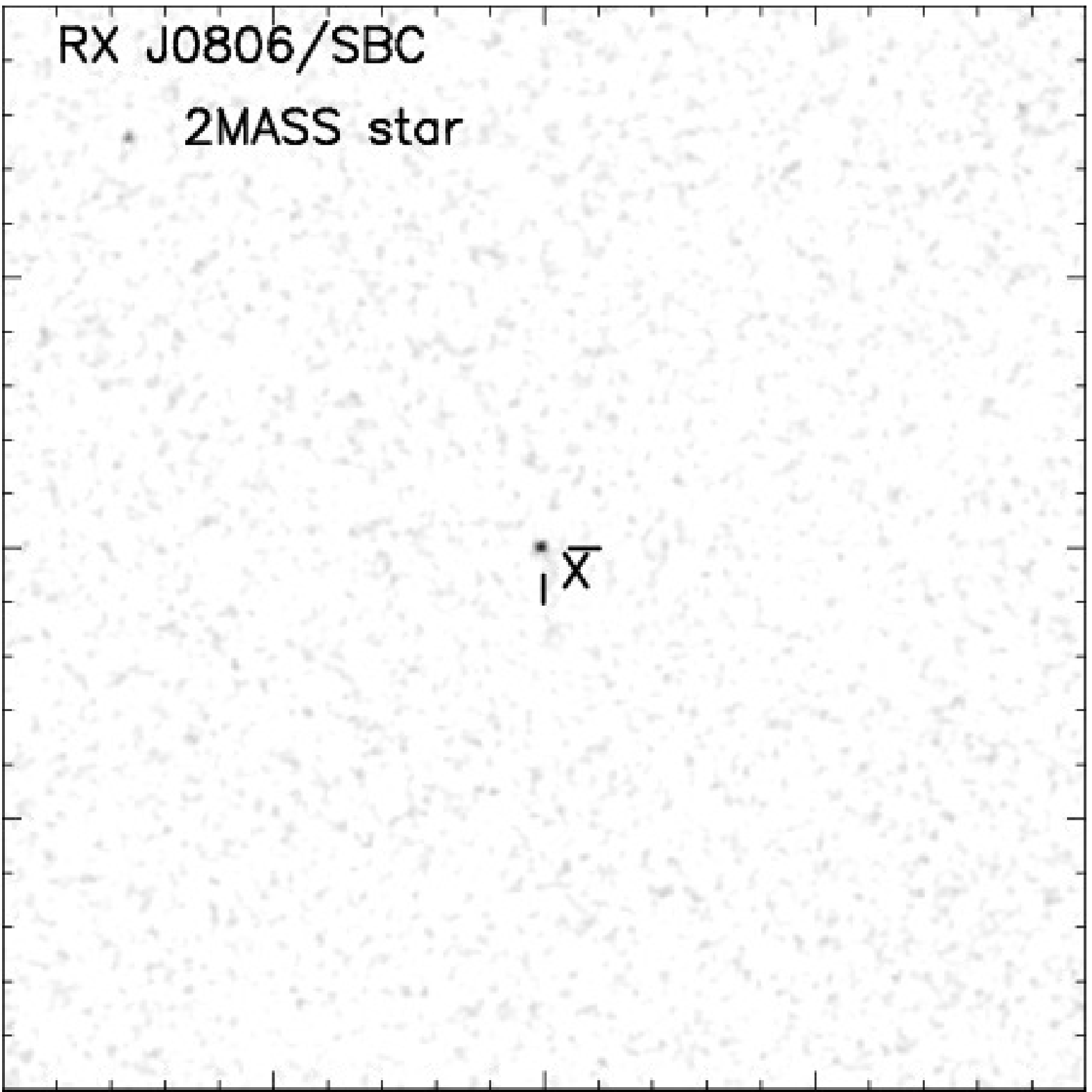}}
\centerline{\includegraphics[width=0.25\textwidth]{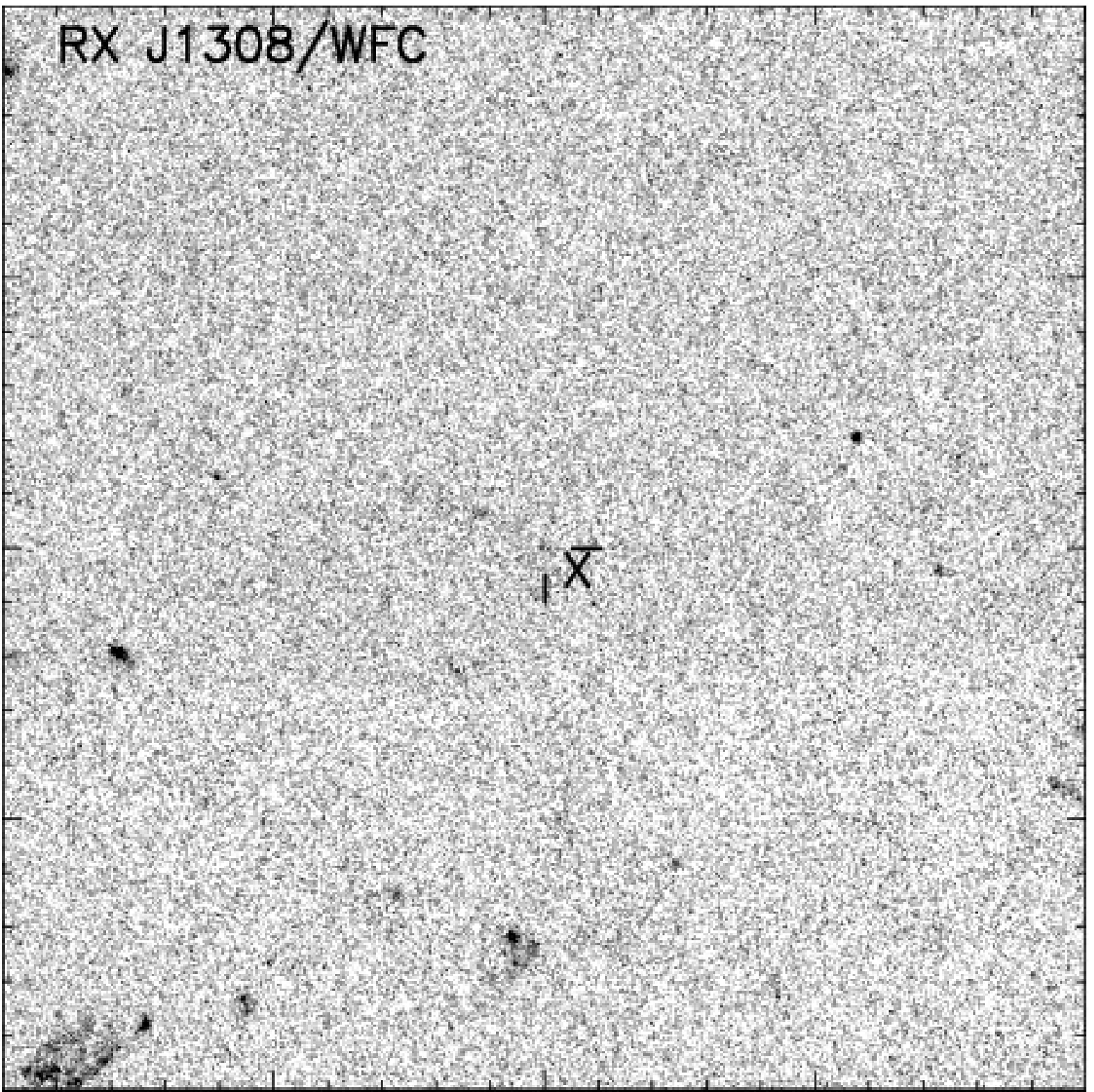}%
\includegraphics[width=0.25\textwidth]{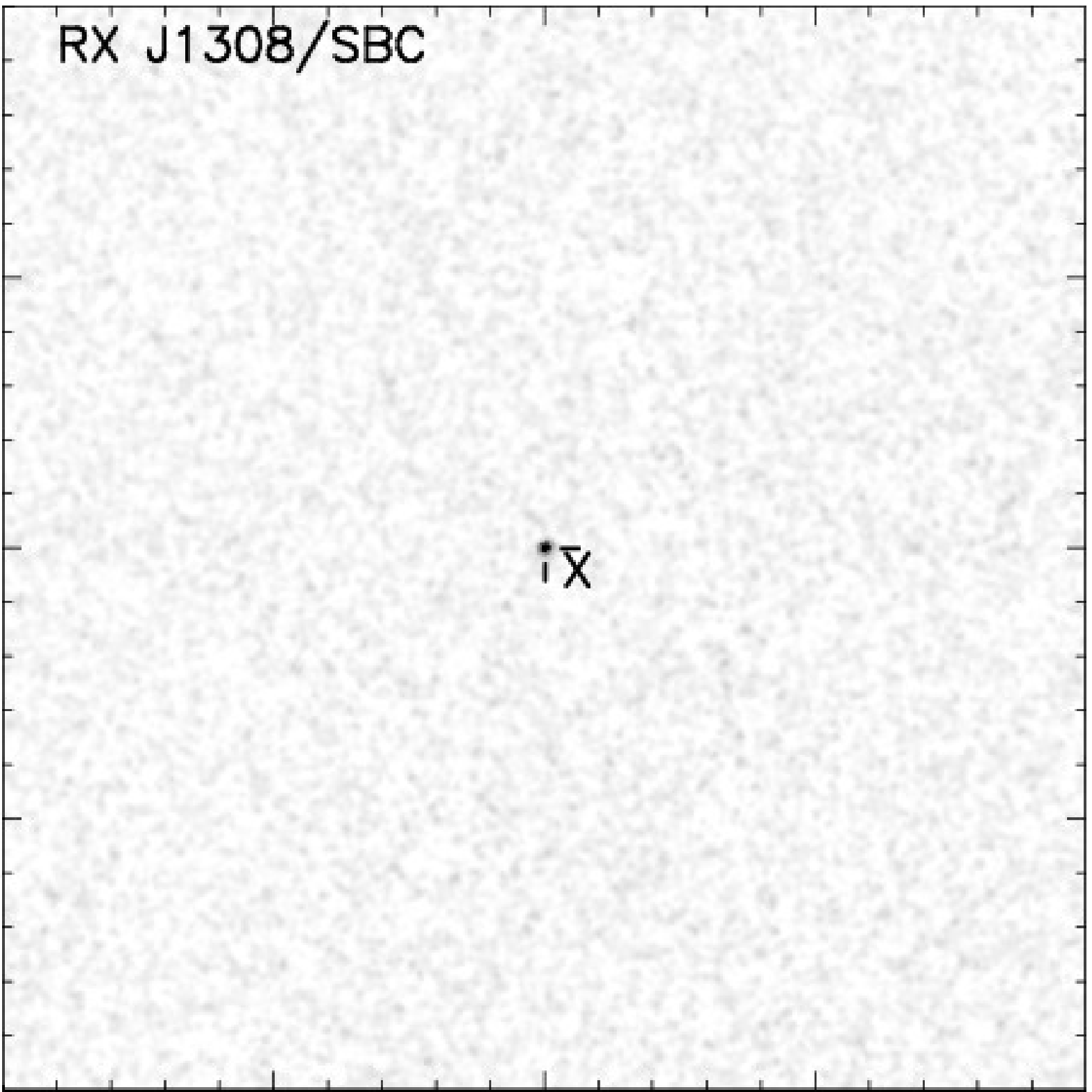}\includegraphics[width=0.25\textwidth]{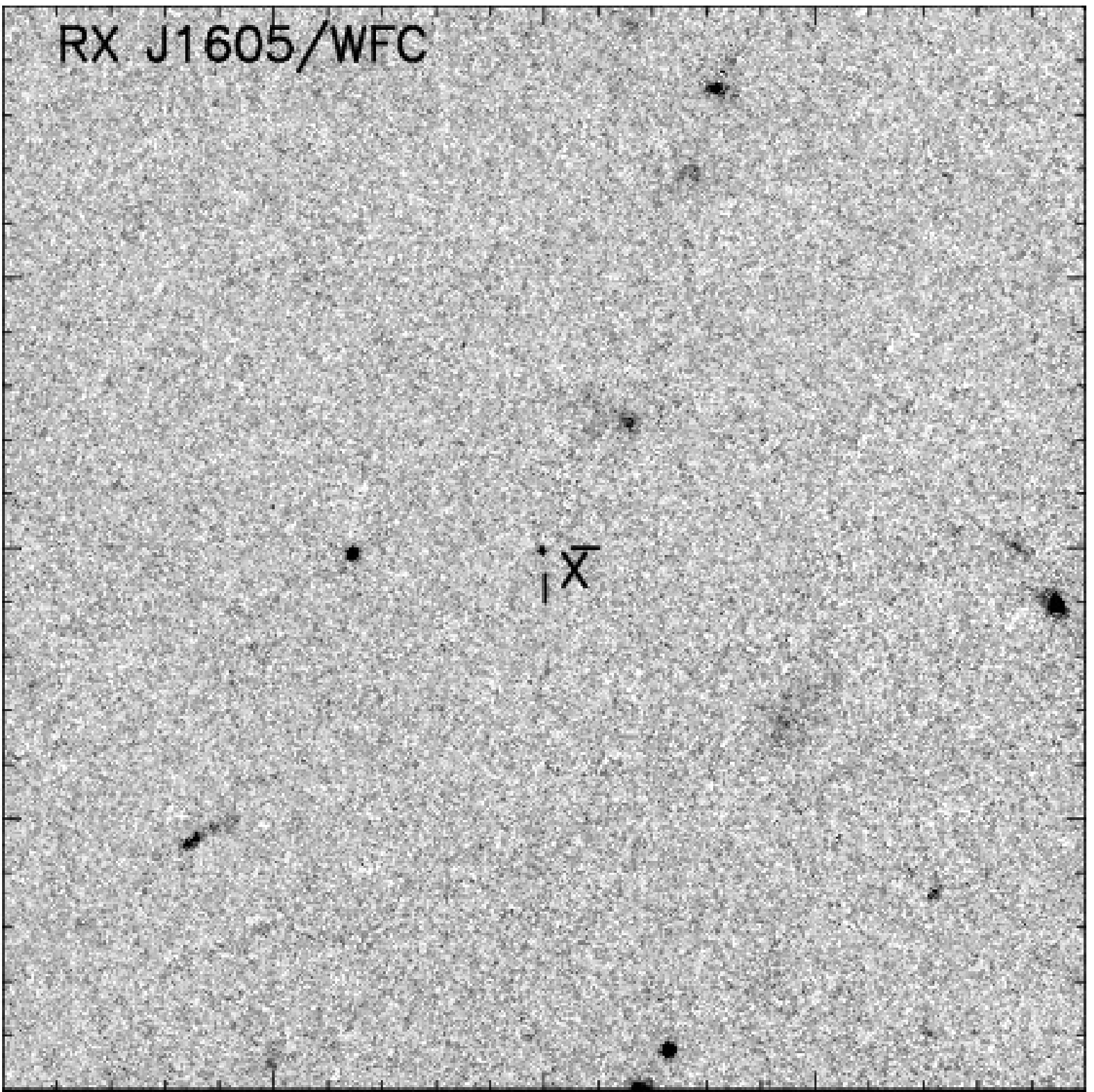}%
\includegraphics[width=0.25\textwidth]{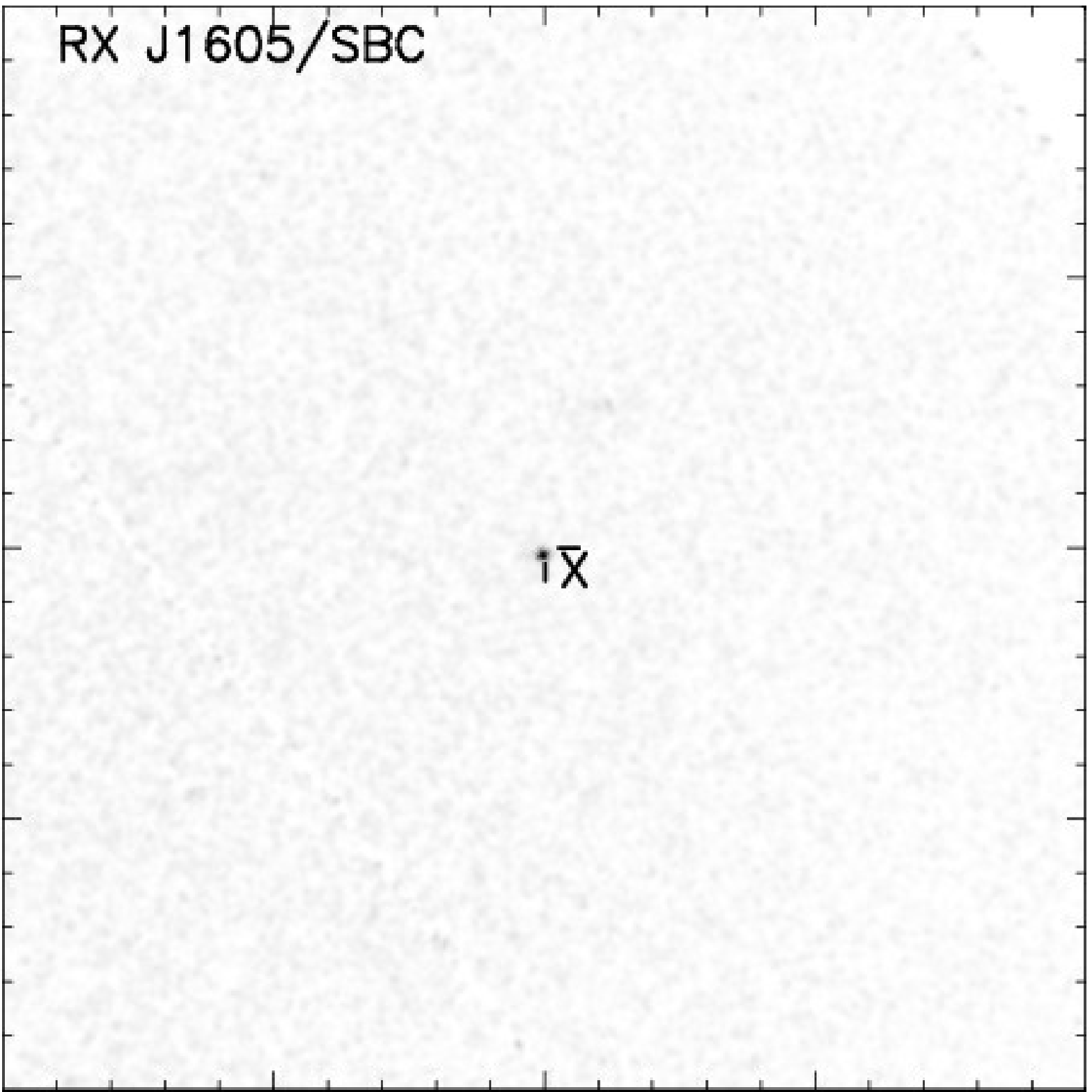}}
{\includegraphics[width=0.25\textwidth]{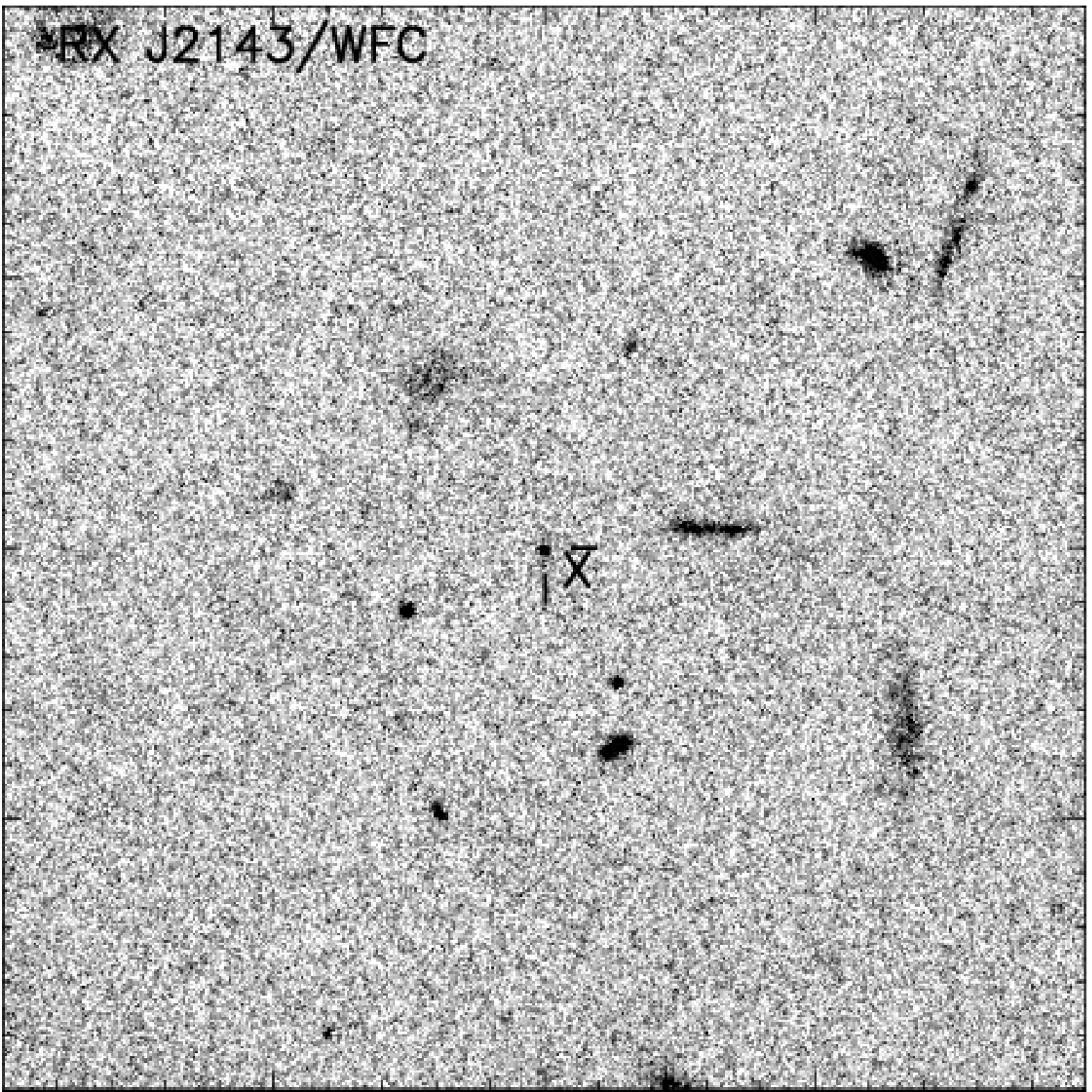}%
\includegraphics[width=0.25\textwidth]{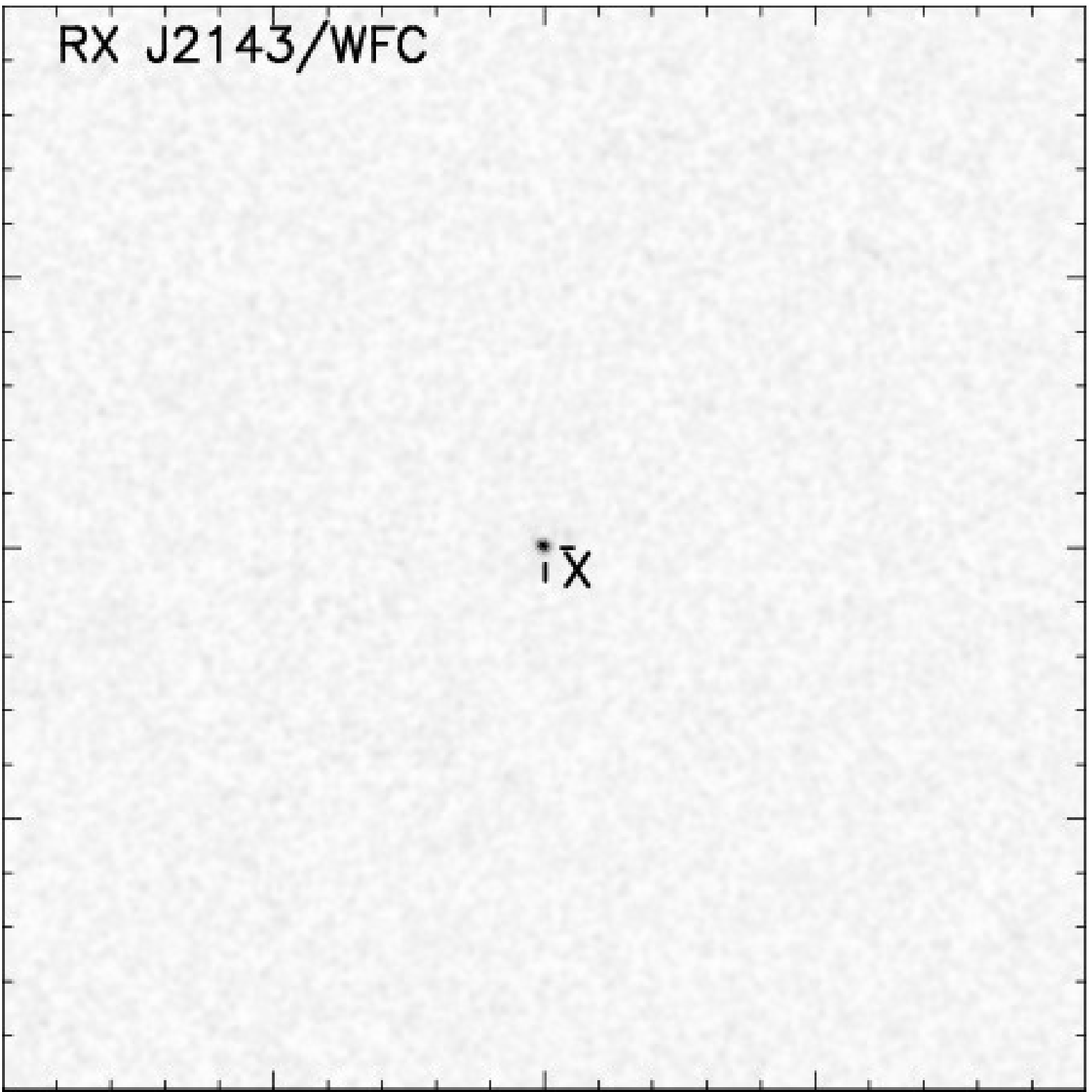}}
\caption{ACS images of the fields around \rxjb\ (top left), \RXJ\ (top
  right), \rbs\ (center left), \rxjvk\ (center right), and
  \rbsb\ (bottom).  For each source we plot WFC/F475W and SBC/F140LP
  images.  The optical counterparts are labeled with ``X'''s; their
  positions agree to better than $0\farcs05$ with bright sources in
  the SBC/F140LP images.  For \rxjb, selected reference sources from
  \citet*{hpm99} are labeled.  The positions of objects ``A'' and a
  galaxy to the south-east match between the optical and UV images.
  For \RXJ, the bright star to the north-east
  (2MASS~J08062407$-$412223.1) is present on both images with the same
  position. 
North is up, and east is to the left,
  and the images are $20\arcsec \times 20\arcsec$.  The SBC images
  have been smoothed with a Gaussian kernel with $\sigma=2$\,pixels.}
\label{fig:images}
\end{figure*}

For all of the WFC images, we then registered the astrometry to 2MASS
\citep{2mass}.  We fit for a shift only, leaving the plate-scale and
rotation as set by the drizzling software.  In the case of \RXJ, we
used 49 stars, restricted to being on the same detector (WFC1) as the
presumed counterpart.  For the other images, many fewer stars were
available, and we used between 1 and 4 (all unsaturated and with
stellar profiles).  From the star-to-star variations, we infer that
our absolute astrometry is accurate to $\sim\!0\farcs2$ only, but this
suffices to show that the positions of the identified counterparts
agreed with published X-ray positions (and generally with previous
identifications; see Sect.~\ref{sec:cpt}).  We defer more accurate
relative and absolute astrometry to a later paper.

\begin{deluxetable*}{l c c c c l}
\tablewidth{0pt}
\tabletypesize{\scriptsize}
\tablecaption{X-ray Blackbodies for the INSs\label{tab:xray}}
\tablehead{
\colhead{RX~J} & 
\colhead{$N_{\rm H}$} & \colhead{$kT$} & \colhead{$R/d$} &
\colhead{Ref.} & \colhead{Notes} \\
& \colhead{($\times 10^{20}\,{\rm cm}^{-2}$)} & \colhead{(eV)} &
\colhead{(${\rm km\,kpc}^{-1}$)} & 
}
\startdata
0420.0$-$5022 & $2.02\pm0.71$ & $\phn45.0\pm2.6$ & $12.2\pm0.2$ &
1 & Normalization is from a blackbody fit. \\
0720.0$-$3125\tablenotemark{a} 
& $1.04\pm0.02$ & $\phn88.4\pm0.4$ & $15.7\pm0.2$ & 2 &
From 2002-Nov, using \xmm/EPIC\\
0806.4$-$4123 & \phn$1.7\pm0.2$\phn & $\phn87.2\pm1.1$ & $\phn9.6\pm0.6$ &
3 & Blackbody with 2 Gaussians\\
1308.6+2127   & $\phn1.8\pm0.2\phn$ & $\phn\phantom{.}102\pm2\phantom{.}\phn$
& $\phn\phn\phantom{.}8\pm2\phantom{.}\phn$ & 4 &
Normalization recomputed from \cxo/LETG\\
1605.3+3249
& $0.98\pm0.19$ & $\phn92.6\pm0.8$ & $12.0\pm0.6$ & 5 & Blackbody
with 1 Gaussian in wavelength, \xmm/EPIC\\
1856.5$-$3754 & $0.95\pm0.03$ & $\phn63.5\pm0.2$ & $36.7\pm0.8$ &
6 & \\
2143.0+0654   & $2.28\pm0.09$ & $104.0\pm0.4$ & $\phn6.2\pm0.1$ & 7\\
\enddata
\tablenotetext{a}{The X-ray spectrum of \rxjk\ evolves
 \citep{dvvmv04}; while little if any evolution is present in the optical
 flux \citep{kvka07}, we attempted to find an X-ray spectrum close in
 time to the \hst\ observations from \citet{kvkm+03} which were taken
 between 2001-July and 2002-Feb.}
\tablecomments{The normalization and temperature are as viewed by an observer at
 infinity.  Where multiple spectral fits are given in a paper, we try
to be specific about which we have selected.  While the best-fit
spectra in many cases include absorption components, we only list the
relevant blackbody parameters here.
}
\tablerefs{1: \citet{hmz+04}; 2: \citet{hhv+09};
 3: \citet{kvk09}; 4: \citet{shhm07}; 5: \citet{vkkd+04}; 6:
 \citet{bhn+03}; 7: \citet{kvk09b}.}
\end{deluxetable*}

\subsection{Photometry}
We carried out aperture photometry of the final drizzled images using
standard procedures in \texttt{IRAF}.  For the WFC data, we used an
aperture of $0\farcs2$, and measured the sky between radii of
$0\farcs4$ and $0\farcs8$ (for comparison, the effective FWHM in our
drizzled images is $0\farcs1$).  Based on \citet{sjb+05}, the aperture
correction from $0\farcs2$ to infinite radius for the
F475W filter is $0.183$\,mag.  There is a small correction to our
aperture correction, since some of the light from the stellar PSF will
appear in the sky annulus.  This will bias the sky value upward and
hence make the source appear fainter than it is.  The fraction of the
stellar light in the sky annulus is 0.04\,mag, based on
\citet{sjb+05}.  Given the ratio between the areas of the source
aperture and the sky annulus, the implied bias to our photometry is
0.003\,mag, considerably less than our photometric uncertainty.  For
F606W, we proceeded similarly, except since we could not use the CTE
correction of \citet{ab10} on those data, we applied an empirical
correction of 0.034\,mag to the photometry, as inferred from the work
of \citet{clkp+09}.

For the SBC data our procedure was similar.  We estimated the sky
background by computing a mean of the data between radii of $1\farcs2$
and $2\farcs4$; here, we found it necessary to turn off the default
clipping of outliers, as this removed valid data: the sky is so dark
that pixels contain only one or two counts, leading to a highly
non-Gaussian distribution of sky values.  Source photometry was done
within a radius of $0\farcs24$, and we used an aperture correction of
$0.33$\,mag, based on the work of
\citet{pbmd03}\footnote{\citet{pbmd03} was specifically concerned with
 the STIS FUV detector.  However the ACS/SBC is a copy of that
 detector, and the aperture corrections should be similar (accounting
 for slightly different plate-scales).  This approach was recommended
 to us by the STScI help desk.  Charge scattering in the SBC detector
 creates a halo extending out to roughly $1\arcsec$ that contains
 about 20\% of the light, according to the ACS Data Handbook
 (\url{http://www.stsci.edu/hst/acs/documents/handbooks/DataHandbookv5/acs\_Ch62.html\#111354}),
 accounting for the large magnitude of the aperture correction.
 While our targets do not have enough counts for very accurate
 profiles out to these radii, a comparison of all of the stars agrees
 reasonably well with the aperture corrections from \citet{pbmd03}
 for a wavelength of 1500\,\AA.}.  Given the large aperture correction
and the possibility of scattered light beyond $1\arcsec$, we include a
5\% systematic uncertainty on our SBC photometry based on the scatter
between the PSFs of the different objects.  The final photometry are given in
Table~\ref{tab:HST_obs}.  All photometry is in the STMAG system, which
is defined so that a source of constant $F_{\lambda}$ has a constant
magnitude: ${\rm STMAG}=-2.5\log_{10} F_{\lambda}-21.1$ with
$F_{\lambda}$ in ${\rm erg\,cm^{-2}\,s^{-1}\,\AA^{-1}}$.

\section{Counterpart Identification and Spectral Fitting}
\label{sec:cpt}
We show the images for all sources in Figure~\ref{fig:images}.  As
discussed above, in all cases there are faint optical sources at the
positions of the ultraviolet sources, whose positions also agree with
the X-ray positions.  Thus, our identifications are secure.  Below, we
first compare our results with previous work, and then discuss the
ultraviolet and optical fluxes, both on their own and in relation to
X-ray measurements.

\paragraph{\rxjb} \citet{hmz+04} identified a possible counterpart,
which is roughly consistent in position.  However, at $B=26.6\pm0.3$,
it is almost a magnitude brighter than our counterpart (using a
correction from STMAG to Vega-based $B$-band of $m_{\rm
  F475W}-B=-0.26$, computed using \texttt{synphot} for a
Rayleigh-Jeans spectrum; this should be accurate to better than 1\%).
It may be that the object identified by \citet{hmz+04} was a blend of
the true counterpart and another object of similar brightness about
$0\farcs5$ to the North (labeled ``Y'' in Figure~\ref{fig:images},
although it is hard to see in that image).  Our photometry is much
more consistent with the object proposed by \citet{mmh+09}.

\paragraph{\rbs} \citet*{kkvk02} identified the same counterpart as we
did.  Using STIS with no filter, which gives a broad band centered
near 4200\,\AA, they measured $m_{\rm 50CCD}=28.56\pm0.13$.  Given
$m_{\rm 50CCD}-m_{\rm F475W}=0.36\,$mag (again, using
\texttt{synphot}), this is consistent with our photometry.

\paragraph{\rxjvk} While the counterpart is secure based on both
colors and proper motion \citep{kkvk03,msh+05,zdlmt06}, there is some
question regarding the optical photometry, with \citet{zdlmt06}
arguing the it cannot be fit with a single power law.  We do not
confirm this: with our re-analysis of the F606W data, the space-based
photometry is consistent with a power law (see below).  For the
ground-based observations of \citet{msh+05}, only the $R$-band point
is consistent with what we measure ($B=27.22\pm0.10$ and
$R=26.9\pm0.14$, and, using {\tt synphot}, $m_{\rm F475W}-B=
-0.32$\,mag and $m_{\rm F606W}-R=0.23$).

\paragraph{\rbsb} We identify the same object as that found by
\citet{zmt+08} and \citet{sek+09}.  These authors each present
$B$-band measurements, of $B=27.4\pm0.2$ and $26.96\pm0.20$,
respectively.  These are only marginally consistent; our measurement
is more consistent with the latter.  \citet{mzt+11} set an upper limit
of $R\gtrsim 27$, which is a bit fainter than our expected value of
$R\approx 26.6$ (from {\tt synphot} we find $m_{\rm
  475W}-R=-0.26$\,mag).  These discrepancies may be from calibration
uncertainties (\citealt{mzt+11} found the two $B$-band measurements to
be consistent given uncertainties in atmospheric extinction
correction), but at least for this source we cannot yet rule out
variability or spectral curvature.

Overall, our identifications our consistent with prior ones, but some
of the fluxes are discrepant.  We do not undertake here to resolve
this, but note that there is no evidence for variability or
non-power-law optical-UV SEDs in the \hst\ data we use below.  We
suspect that the discrepancies reflect mostly the more than usual care
required to obtain reliable ground-based measurements for very faint
objects (see, e.g., the discussion for \rxjk\ in \citealt{mzh03}).

\subsection{Power-law Fits}
As in \citet{vkk01} and similar papers, we tried fitting single
power-laws to just the optical/UV data for all 7 INSs; we included the
space-based photometry for \rxjw\  \citep{vkk01} and for \rxjk\
 \citep{kvkm+03}.  Detailed fitting of \hst\ data for the INSs
presents some difficulties, since the zero-point fluxes were
determined for a spectrum that is flat in $F_\lambda$, while the INSs
have much steeper spectra.  Given the often very wide bandpasses used,
this leads to substantial changes in effective wavelength.

\begin{figure*}
\centerline{\includegraphics[height=0.8\textheight]{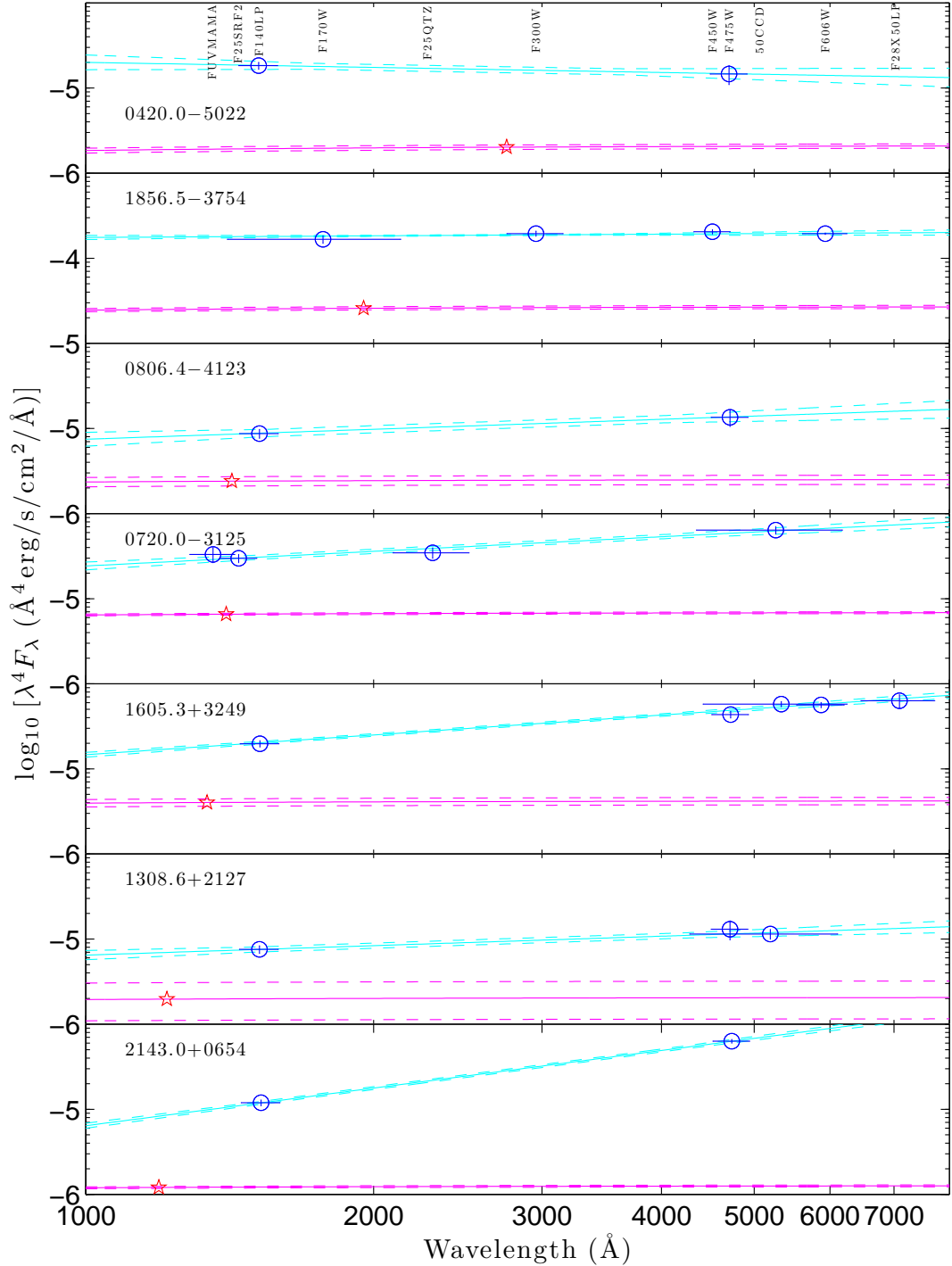}}
\caption{Optical/UV SEDs for the 7 INSs.  For each object (sorted by
 increasing $kT$ from top to bottom), we plot $\lambda^4 F_\lambda$
 versus wavelength; a Rayleigh-Jeans tail would be flat.  The
 best-fit power-laws with $\pm1\sigma$ uncertainties are shown by the
 cyan lines.  The extrapolations of the X-ray blackbodies with
 $\pm1\sigma$ uncertainties are shown by the magenta lines.  The
 fluxes have been corrected for absorption using the nominal values
 of $N_{\rm H}$ from Table~\ref{tab:xray}.  The red stars show the
 wavelength of $10kT$.  The different filters are
 labeled at the bottom. 
}\label{fig:sed}
\end{figure*}

To circumvent these difficulties, we relied on \texttt{synphot} to
compute the expected STMAG as a function of extinction $A_V$ and
power-law index $\alpha$ (with $F_\lambda \propto \lambda^{-\alpha}$,
where a Rayleigh-Jeans spectrum would have $\alpha=4$) on a fine grid,
with steps 0.005\,mag in $A_V$ and 0.025 in $\alpha$.  Then, for each
object, we fitted a power law, optimizing for $\alpha$ and
normalization, but keeping the reddening $A_V$ fixed to the value
implied by the column density $N_{\rm H}$ determined from fits to the
X-ray spectrum: $A_V=N_{\rm H}/\expnt{1.79}{21}\,{\rm cm}^{-2}$
(\citealt{ps95}; note that this relation varies for different sight
lines, but the extinction is small enough that the variations do not
influence our results qualitatively, as we show below).  We use the
extinction curve of \citet*{ccm89}, implemented as \texttt{gal3} in
\texttt{synphot}, and we assume $E(B-V)=A_V/3.1$.  We note that we
cannot fit uniquely for $A_V$ based only on our data, as the amount of
reddening is highly degenerate with the power-law index.  The exact
value of $N_{\rm H}$ (and hence $A_V$) that we used depends somewhat
on the calibration of the X-ray data and on the assumed spectral
model, and could change in the future.  We discuss this explicitly
below.

The results of our power-law fits are shown in with
Table~\ref{tab:plfit} and Figure~\ref{fig:sed}.  Our uncertainties and
$\chi^2$ values include a 5\% systematic uncertainty for the F140LP
data, owing to uncertainties in the aperture correction.  For those
objects with only two measurements we are left with a fully
constrained fit (0 degrees-of-freedom); for the others, $\chi^2$ is
consistent with a good fit.

The precise results depend slightly on our choice of extinction, but
the results do not change qualitatively, since the extinctions
inferred from the X-ray column densities are small ($A_V\lesssim
0.1\,$mag, and thus $A_{1400}\lesssim0.3\,$mag).  Indeed, given the
quoted uncertainties on $N_{\rm H}$, the changes are nearly
negligible.  Somewhat larger systematic changes are possible from
changes in the reddening law, the ratio of $N_{\rm H}$ to $A_V$, or
the assumed X-ray spectral model.  Therefore, in Table~\ref{tab:plfit}
we also give $d\alpha/dA_V$: the change in the power-law slope with
changes in extinction (see also \citealt{vkk01}).  We note, though,
that even a 50\% change in $A_V$ would generally change $\alpha$ by
less than~0.1.

\subsection{Comparison with X-ray Spectra}
One of the main purposes of our work is to establish reliable
estimates of the ``optical excess,'' the factor by which the
optical/UV flux of the INS exceed the extrapolation of the X-ray
blackbodies.  To do this, we have collected the best-fit X-ray spectra
for all 7 INSs in Table~\ref{tab:xray}.  In most cases the best-fit
spectra are not purely blackbodies, but for the purposes of
extrapolation the blackbody component is sufficient.  We note that the
spectrum of \rxjk\ changed during 2003 \citep{dvvmv04}.  While there
is no evidence that the optical flux of \rxjk\ changed during that
time \citep*{kvka07}, the two different X-ray spectra differ by a
factor of 1.54 from each other, thus introducing a systematic
uncertainty into our inferred optical/UV excess.  It does not,
however, affect the shape of the excess, as the \hst\ data are well
into the Rayleigh-Jeans tail and uncertainties in the extinction are
small.

For all of the objects we determine the excess relative to expected
fluxes calculated by passing the extrapolated, reddened X-ray spectrum
through the \hst\ filter response curves, using \texttt{synphot}.
This avoids issues of uncertainties in the effective wavelengths or
zero-point flux that come from using spectra that differ significantly
in slope from the calibration spectra.  We determined formal
uncertainties by repeating the above for values that differ by
1-$\sigma$ in all parameters from the best-fit spectrum, taking as
1-$\sigma$ uncertainties on the extrapolated magnitudes the maximum
and minimum of these variations.  

We list our results in Tables~\ref{tab:extrap} and \ref{tab:extrap2}.  For completeness, we
include expected values for instrument/filter/object combinations
beyond those used here, and also give reference photometry for an
unabsorbed $10^6\,$K blackbody as well as approximate values for the
wavelength-dependent extinction $A_\lambda/A_V$ (see \citealt{vkk01}
for an extended discussion).  The extinction values actually depend
slightly on the value of the overall extinction $A_V$, since this
changes the shape of the spectrum and hence the integral over the
filter response.  The effect is largest for the widest filters, like
STIS/50CCD.  However, for the range of extinction values considered
here ($N_{\rm H}=\expnt{(1-4)}{20}\,{\rm cm}^{-2}$ corresponds to
$A_V\approx 0.06-0.22$), the changes are less than 1\% in
$A_\lambda/A_V$.

Along with the more easily assessed uncertainties on $kT$ and $(R/d)$,
a possibly significant systematic uncertainty in our model is
variations in extinction.  Similar to the analysis above, we explore
this issue by considering a $\pm 50$\% range on $A_V$ around the
nominal value.  This results in uncertainties on the extrapolated
photometry ranging from 0.09\,mag (for the shortest wavelengths) to
0.03\,mag (for the longest), based on a fiducial $N_{\rm
  H}=10^{20}\,{\rm cm}^{-2}$.  For other values of $N_{\rm H}$, the
uncertainties are proportionate.  Generally, these additional
uncertainties are smaller than the photometric uncertainties, and will
not affect our results.  The measured excesses are given in
Table~\ref{tab:plfit} and Figure~\ref{fig:sed}.

\begin{deluxetable*}{l c c c c c}
\tablewidth{0pt}
\tablecaption{Predicted \hst\ UV Photometry from Extrapolated X-ray Blackbodies\label{tab:extrap}}
\tablehead{
\colhead{RX~J} & \multicolumn{5}{c}{Extrapolated STMAG} \\ \cline{2-6}
& \colhead{STIS/FUV} & \colhead{STIS/FUV} & \colhead{ACS/SBC} &
\colhead{WFPC2} & \colhead{STIS/NUV}  \\
& & \colhead{F25SrF$_2$} & \colhead{F140LP} & \colhead{F170W} &
\colhead{F25Qtz}  
}
\startdata
0420.0$-$5022 & $24.87\pm0.07$ & $25.10\pm0.07$ & $25.30\pm0.07$ & $25.95\pm0.07$ & $27.03\pm0.07$ \\
0720.0$-$3125 & $23.36\pm0.03$ & $23.61\pm0.03$ & $23.82\pm0.03$ & $24.48\pm0.03$ & $25.57\pm0.03$ \\
0806.4$-$4123 & $24.57\pm0.14$ & $24.81\pm0.14$ & $25.01\pm0.14$ & $25.67\pm0.14$ & $26.76\pm0.14$ \\
1308.6+2127   & $24.79\pm0.55$ & $25.03\pm0.55$ & $25.24\pm0.55$ & $25.89\pm0.55$ & $26.99\pm0.55$ \\
1605.3+3249   & $23.88\pm0.11$ & $24.13\pm0.11$ & $24.34\pm0.11$ & $25.00\pm0.11$ & $26.09\pm0.11$ \\
1856.5$-$3754 & $21.88\pm0.05$ & $22.13\pm0.05$ & $22.34\pm0.05$ & $23.00\pm0.05$ & $24.09\pm0.05$ \\
2143.0+0654   & $25.41\pm0.03$ & $25.64\pm0.03$ & $25.84\pm0.03$ & $26.50\pm0.03$ & $27.59\pm0.03$ \\
\hline
$10^6$\,K\tablenotemark{a} & 19.81 & 20.08 & 20.29 & 20.96 & 22.05  \\
$\lambda_{\rm eff}$\tablenotemark{a} (\AA) & 1355 & 1442 & 1517 & 1769 & 2283  \\
$A_\lambda/A_V$\tablenotemark{a} & 3.11 & 2.84 & 2.71 & 2.65 & 2.58 \\
\enddata
\tablecomments{We give the instrument name and filter for all
 UV observations that we consider; not all objects were observed with
 all combinations, and see Table~\ref{tab:extrap2} for the optical observations.
 The 1-$\sigma$ uncertainties are based on the uncertainties in the
 X-ray spectra (Table~\ref{tab:xray}) and do not include any other
 systematic terms.}
\tablenotetext{a}{The STMAGs for a $10^6$\,K (86\,eV) blackbody with
 $A_V=0$ and normalized to $m_{\rm F450W}=25$.  The wavelengths given are approximate effective
 wavelengths for the instrument/filter and a typical INS spectrum. Wavelength-dependent extinction $A_\lambda/A_V$ for a
 $10^6$\,K (86\,eV) blackbody, calculated for $A_V=0.05$.  Effective
 wavelengths and extinctions follow the definition of \citet{vkk01}.}
\end{deluxetable*}
\setlength{\tabcolsep}{5pt}
\tabletypesize{\footnotesize}

\begin{deluxetable*}{l  c c c c c c c}
\tablewidth{0pt}
\tablecaption{Predicted \hst\ Optical Photometry from Extrapolated X-ray Blackbodies\label{tab:extrap2}}
\tablehead{
\colhead{RX~J} & \multicolumn{7}{c}{Extrapolated STMAG} \\ \cline{2-8}
& \colhead{WFPC2} &
\colhead{WFPC2} & \colhead{ACS/WFC} & \colhead{STIS/CCD} &  \colhead{ACS/WFC}&
\colhead{WFPC2} & \colhead{STIS/CCD} \\
& \colhead{F300W} & \colhead{F450W}
& \colhead{F475W} & \colhead{50CCD} & \colhead{F606W}& \colhead{F606W} & \colhead{F28x50LP} 
}
\startdata
0420.0$-$5022 &  $28.06\pm0.07$ & $29.81\pm0.06$ & $29.98\pm0.06$ & $30.35\pm0.07$ & $30.88\pm0.06$ & $30.95\pm0.06$ & $31.66\pm0.06$ \\
0720.0$-$3125 &  $26.64\pm0.03$ & $28.44\pm0.03$ & $28.62\pm0.03$ & $28.96\pm0.03$ & $29.54\pm0.03$ & $29.60\pm0.03$ & $30.32\pm0.03$ \\
0806.4$-$4123 &  $27.81\pm0.14$ & $29.59\pm0.14$ & $29.76\pm0.14$ & $30.11\pm0.14$ & $30.67\pm0.14$ & $30.73\pm0.14$ & $31.45\pm0.14$ \\
1308.6+2127   &  $28.03\pm0.55$ & $29.81\pm0.55$ & $29.98\pm0.55$ & $30.33\pm0.55$ & $30.89\pm0.55$ & $30.95\pm0.55$ & $31.67\pm0.55$ \\
1605.3+3249   &  $27.17\pm0.11$ & $28.97\pm0.11$ & $29.15\pm0.11$ & $29.49\pm0.11$ & $30.07\pm0.11$ & $30.13\pm0.11$ & $30.85\pm0.11$ \\
1856.5$-$3754 &  $25.16\pm0.05$ & $26.96\pm0.05$ & $27.13\pm0.05$ & $27.48\pm0.05$ & $28.05\pm0.05$ & $28.12\pm0.05$ & $28.84\pm0.05$ \\
2143.0+0654   &  $28.62\pm0.03$ & $30.37\pm0.03$ & $30.54\pm0.03$ & $30.91\pm0.03$ & $31.44\pm0.03$ & $31.51\pm0.03$ & $32.22\pm0.03$ \\
\hline
$10^6$\,K\tablenotemark{a} &
23.16 & 25.00 & 25.18 & 25.50 & 26.11 & 26.18 & 26.91 \\
$\lambda_{\rm eff}$\tablenotemark{a} (\AA) & 2955 &
4520 & 4709 & 5066 & 5844 & 5932 & 7025 \\
$A_\lambda/A_V$\tablenotemark{a} &
1.93 & 1.30 & 1.24 & 1.58 & 0.97 & 0.95 & 0.79\\
\enddata
\tablecomments{We give the instrument name and filter for all
 optical observations that we consider; not all objects were observed with
 all combinations, and see Table~\ref{tab:extrap} for the optical observations.  
 The 1-$\sigma$ uncertainties are based on the uncertainties in the
 X-ray spectra (Table~\ref{tab:xray}) and do not include any other
 systematic terms.}
\tablenotetext{a}{The STMAGs for a $10^6$\,K (86\,eV) blackbody with
 $A_V=0$ and normalized to $m_{\rm F450W}=25$.  The wavelengths given are approximate effective
 wavelengths for the instrument/filter and a typical INS spectrum. Wavelength-dependent extinction $A_\lambda/A_V$ for a
 $10^6$\,K (86\,eV) blackbody, calculated for $A_V=0.05$.  Effective
 wavelengths and extinctions follow the definition of \citet{vkk01}.}
\end{deluxetable*}
\setlength{\tabcolsep}{5pt}
\tabletypesize{\footnotesize}

\section{Discussion \& Conclusions}
\label{sec:disc}
\subsection{Comparing the INSs to Each Other}
It is clear that there is a wide variation in both the normalization
and slope of the optical/UV power-laws for the INSs.  Comparing  to
previous work, we confirm the conclusion of \citet{vkk01} that
\rxjw\ has a nearly thermal spectrum, close to $F_\lambda \propto
\lambda^{-4}$, while the spectrum of \rxjk\ is somewhat flatter,
consistent with what was found by \citet{kvkm+03} and \citet{mzh03}.
The other objects extend the range, with \rxjb\ also appearing thermal
and \rbsb, with $\alpha=2.5$, having the flattest, least
Rayleigh-Jeans-like spectrum (as well as the highest excess; see
below).  No source has a slope that is significantly steeper than
Rayleigh-Jeans.

Comparing the emission to the X-ray spectrum, there is a similarly
wide variation in both the amount and slope of the optical/UV excess:
most sources are not consistent with the constant excess of \rxjw, and
the amount of the excess varies significantly, although no source has
measurements below its X-ray extrapolation.  At 1500\,\AA\ the range
is from about 3.5 to almost 10, with statistically significant
variation ($\chi^2=134$ for 6 DOF) even when excluding
\rbsb\ ($\chi^2=46$ for 5 DOF).  At longer wavelengths the variation
is even larger, going from 6 to 50 at 4700\,\AA.  Here, \rbsb\ is
clearly very different from the rest, but again when excluding it
there is still significant variation ($\chi^2=21$ for 5 DOF).  These
are beyond deviations possible from the X-ray uncertainties, and while
X-ray calibration errors, differences in X-ray fitting methodology,
and variability can all contribute to variations in the excess, as
noted above the shape of the excess is robust.  See
Figure~\ref{fig:ox} for a full optical-to-X-ray comparison of
\rxjw\ and \rbsb.

If the X-ray and optical came from different regions on the surface
\citep{br02}, we might expect variations in the excess to correlate
with changes in the X-ray pulsed fraction, as both would be dependent
on geometry: a small X-ray hotspot would give rise to both a large
optical excess and a large pulsed fraction.  Some allowances could be
made for viewing angles, but with 7 sources this would start to
average out.  However, we do not see any such correlation.  As an
example, both \rxjw\ and \rxjb\ have similar optical excesses, but the
pulsed fraction for \rxjw\ is about 1\% \citep{tm07} while that of
\rxjb\ is 13\% \citep{hmz+04}, and \rxjvk\ has a larger excess but a
small pulsed fraction.

\begin{figure}
\plotone{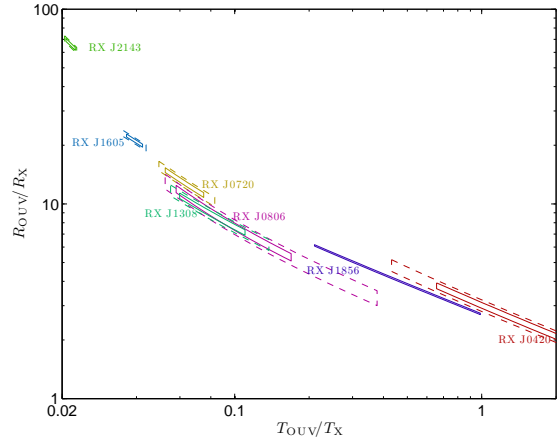}
\caption{Results of fitting blackbodies to only the optical/UV data
  for the INSs.  We show $\pm1\sigma$ (solid) and $\pm2\sigma$
  (dashed) limits on the temperature and normalization $R/d$, where in
  both cases we scale the fitted values by the values measured in
  X-rays (causing the distances to cancel).  Individual sources are
  labeled.  For \rxjw\ and \rxjb, the results are consistent with the
  X-ray temperature at a constant excess, as seen from
  Table~\ref{tab:plfit}.  Other objects have flatter spectra so the
  implied fits are cooler.  For the objects with more than two data
  points and $T_{\rm OUV}<T_{\rm X}$, \rxjk\ has $\chi^2=4.3$ (2 DOF)
  and \rxjvk\ has $\chi^2=5.2$ (3 DOF).  Extrapolating the optical
  fits to higher energies for \rxjb\ and to a lesser extent
  \rxjw\ might not be consistent with the X-ray data.}
\label{fig:bbfit}
\end{figure}

Even so, we can still fit the optical/UV SEDs as part of some
blackbody, with radius and temperature free (Fig.~\ref{fig:bbfit}).
For \rxjw\ this is largely degenerate (since it looks like a
Rayleigh-Jeans), but for other sources the temperature and hence the
radius are constrained.  Some of these (mostly \rxjb) are not likely
based on extrapolations to the X-ray band, as they would exceed the
flux at 100\,eV.  All fits are generally consistent with reduced
$\chi^2\approx 1$, although there are not many degrees of freedom.
This fit gives a normalization of $R/d\approx 400\,{\rm km\,kpc}^{-1}$
for \rbsb.  Since we expect a distance to \rbsb\ of 500--1000\,pc
based on the X-ray spectrum, this would imply a rather large radius if
it is interpreted physically, and may only apply in light of the
scattering model discussed below.  Most of the objects fall on a
single locus with radius increasing as the temperature decreases.
Much of this just comes from keeping a similar excess among the
sources (we would expect $R^2\propto 1/T$), but it is clear that
\rxjvk\ and \rbsb\ are inconsistent with the other objects.

\begin{figure}
\plotone{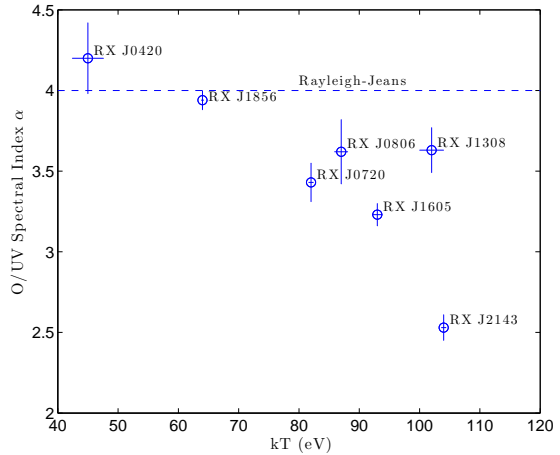}
\caption{Optical/UV spectral index ($\alpha$ from
 Table~\ref{tab:plfit}) versus X-ray temperature $kT$ (from
 Table~\ref{tab:xray}) for the INSs.  The horizontal line shows a
 slope of 4, to be expected for a Rayleigh-Jeans tail.
}
\label{fig:ktslope}
\end{figure}

We saw above that the optical excess does not appear to correlate with
the X-ray pulsed fraction, suggesting that a model like \citet{br02}
is incomplete.  Taking the basic spectral parameters ($kT$ along with
energy of the absorption lines) and the basic rotational parameter
(pulsed fraction, magnetic field, spin-down luminosity) we see no
clear trends with the optical/UV properties that we have measured
here.  The one possible exception is in relating the spectral index to
$kT$: in Figure~\ref{fig:ktslope} it appears that the hotter objects
have smaller spectral indices.  Much of this correlation comes from
\rbsb, but even without it the correlation appears somewhat
significant: Spearman's rank correlation coefficient
\citep[][p.\ 641]{numrec} for all objects is $-0.75$ with a
null-hypothesis probability of 5\%, and this drops to $-0.6$ (21\%
null hypothesis probability) when \rbsb\ is excluded.  For the small
number of objects considered here, this is reasonable but not
definitive.

Such a correlation may imply that the SEDs of all of the INSs deviate
from a Rayleigh-Jeans tail, but that the energy at which they do it
depends on their temperatures.  For the cool \rxjw\ and \rxjb, the
optical/UV window still seems thermal.  But for the hotter \rbs\ and
\rbsb, the thermal portion would be lost shortward of 1000\,\AA.  The
shift does not appear linear, as measuring the excess at a wavelength
that scales with $1/kT$ does not improve the agreement among objects.

In contrast to some previous attempts to measure the optical SED using
both ground- and space-based measurements, we see that all INSs are
well fit by a single power-law.  Further, our results using only
\textit{HST} data are similar to those for \rxjw\ and \rxjk\ that
include ground-based data.  Our conclusions regarding the 5 sources
observed here are relatively insensitive to $\sim$month scale
variability, as in many (but not all) of the cases the optical and UV
observations occurred within weeks of each other, but this is not
always the case.  For instance, \rxjvk\ now has more \hst\ photometry
than \rxjw\ spanning almost 10 years, but still appears smooth.
Whether this implies that the optical/UV is constant or just that we
are unlucky is not clear, but future high-quality measurements should
be able to address this.

\subsection{Comparing the INSs to Pulsars and Magnetars}
Previous modeling of \rxjk\ \citep{kvkm+03} could not distinguish
between a single non-thermal power-law and a combination of
power-laws.  Unfortunately, with our new data it is still hard to make
that distinction, as the wide wavelength gap between the optical and
UV even in the best case (\rxjvk) could easily hide a curved SED.  
Future observations in the 2000--3000\,\AA\ range might help better
constrain such spectra and settle this question.  We note, though,
while the sources with few data-points can accommodate a power-law as
flat as that for \rbsb\ in addition to a steeper, thermal component,
objects with denser coverage like \rxjk\ and \rxjvk\ cannot.

We can set rough upper limits on another power-law by requiring that
any non-thermal contribution match our reddest optical point as well
as be below the X-ray black body at 1.5\,keV, where typically there
are no more counts (Figure~\ref{fig:ox}).  We also require that such a
power-law have a slope such that it is falling in photons per unit
energy toward higher energies, so that it stays hidden
\citep[see][]{kvkm+03}.  Taken together with the parallax for
\rxjw\ \citep{wel+10} and assuming all the INSs have similar radii, we
find non-thermal luminosities integrated over the 2--10\,keV band of
typically $10^{28-30}\,{\rm erg\,s}^{-1}$.  At the upper end we do not
believe that such luminosities could be real, as they would be close
to 100\% of $\dot E$, in contrast to radio pulsars that have
non-thermal luminosities of $\sim 10^{-3} \dot E$ in X-rays
\citep{bt97}.  Even without such an extrapolation our optical
luminosities ($\expnt{(0.3-1.7)}{29}\,{\rm erg\,s}^{-1}$) are
uncomfortably high for a typical magnetospheric origin, since it would
be $>10^{-3}\dot E$ compared to typical values of ratios of
$10^{-7}$--$10^{-6}$ for radio pulsars \citep{zp04}.  However, while
the X-ray emission cannot be powered by $\dot E$ (since X-ray
luminosities are $\gg \dot E$; \citealt{kvk09b}) some amount of the
optical emission could be related, albeit with a different mechanism
than that which usually operates for pulsars.

If the deviations from a thermal SED were from magnetospheric emission
as in pulsars \citep{pwc97,ssl+05,kpzr05}, we might expect a
correlation in the amount of the excess or the optical luminosity with
the spin-down luminosity $\dot E$, as this is what drives the X-ray
power-laws seen in pulsars and may drive the optical too.  Again,
though, we see no such correlation.

The behavior of the INSs also deviates from what is seen in the cases
of other relatively young neutron stars in the optical, namely
magnetars \citep{mereghetti08}. Most magnetars have only been seen in
the infrared (likely due to high extinctions).  The infrared emission
is variable, and may or may not be linked to the X-ray emission.  The
one source with confirmed optical emission is 4U~0142+61
\citep*{hvkk00,hvkk04}.  The optical emission is known to pulse at the
X-ray period \citep{km02,dmh+05}, suggesting that it is likely
magnetospheric in origin \citep*{znt11}.  Importantly for this
comparison, the optical emission is not a single power-law
\citep{hvkk04}.  In contrast, the emission of the INSs is well fit by
a single power-law and is (so far) constant in time.  The flattening
of the SED of \rbsb\ toward longer wavelengths is somewhat suggestive
of the broadband SED of 4U~0142+61, but the other characteristics make
it distinct. 

It may be that the similarities with magnetars arise from a common
origin, which suggests that the optical emission results from resonant
scattering in the neutron stars' magnetospheres \citet{lg06}.
However, qualitative differences between the INSs and magnetars exist:
while the ranges overlap, the magnetars typically have higher values
of magnetic to thermal energy $\hbar e B/m_e c k T$ ($\gtrsim 10^4$ for the magnetars
vs.\ $\approx 3000$ for the INSs), and the higher fields of the
magnetars mean that scattering and pair-production happen much more
quickly.  The optical emission of the INSs could instead be from an
inverted temperature layer (like a chromosphere) that is driven by a
twisted magnetosphere (C.~Thompson 2011, pers.\ comm.).  We note that
for \rbsb, which has the most extreme optical excess, we also see that
the X-ray pulsations have odd harmonics likely indicative of a complex
field geometry \citep{kvk09}.

\subsection{Atmosphere Models for the INSs}
In the context of magnetized atmosphere models for neutron stars, the
amount of optical/UV excess can depend on the magnetic field strength.
Therefore we examined the UV/optical emission predicted by the NSMAX
models of NS X-ray spectra constructed for \texttt{Xspec}
\citep{hpc08}; these models span the range $B=(1-30)\times 10^{12}$\,G
and $kT\sim 20-400$\,eV for a partially ionized hydrogen atmosphere.
We find that, although the magnitude of the flux differs from that of
a blackbody, the wavelength dependence still exhibits a $\lambda^{-4}$
Rayleigh-Jeans behavior.  Deviations from the Rayleigh-Jeans behavior
occur as a result of the proton cyclotron line at (redshifted) $\sim
2500/B_{12}$\,\AA\ (with $B=10^{12}B_{12}$\,G).  For $B\lesssim
4\times 10^{12}$\,G, the wings of the cyclotron line can reproduce the
wavelength-behavior seen in Fig.~\ref{fig:sed}, though not at the
magnitude of \rbsb; however, it is unclear how strongly this behavior
depends on temperature (see Fig.~\ref{fig:ktslope}).  If the
deviations from Rayleigh-Jeans are due to absorption in the wings of
the proton cyclotron line, these spectrally-inferred magnetic fields
are much lower than those inferred from timing \citep{kvk09b}.  A
factor of two increase in the former may be the result of a helium
(rather than hydrogen) atmosphere.  A possible discrepancy between
timing and spectral magnetic fields may exist when considering the
broad X-ray absorption features \citep{haberl07,vkk07,kvk09} but there
it is in the opposite direction (the features are more consistent with
strong fields), and hence easier to understand in terms of
higher-order magnetic moments leading to locally stronger magnetic
fields.  We also note that our models indicate an increasing flux with
increasing wavelength similar in magnitude to that seen in the INSs
(again except for \rbsb) for $kT\sim 100$\,eV and magnetic fields more
similar to what is seen in X-ray timing, $B\approx (2-3)\times
10^{13}$\,G; the cause of this is still under investigation.

\begin{deluxetable*}{l c c c c c c c c}
\tablewidth{0pt}
\tabletypesize{\scriptsize}
\tablecaption{Power-law Fits to \hst\ Photometry\label{tab:plfit}}
\tablehead{
\colhead{RX~J} & \colhead{$A_{V,\rm ref}$} & \colhead{$F_{\lambda}(2500\,{\rm \AA})$} &
\multicolumn{3}{c}{Excess} &\colhead{$\alpha$} &
\colhead{$d\alpha/dA_V$} & \colhead{$\chi^2$/DOF} \\ \cline{4-6}
& \colhead{(mag)} & \colhead{($10^{-18}$\,erg\,cm$^{-2}$\,s$^{-1}$\,\AA$^{-1}$)} &
\colhead{1500\,\AA} & \colhead{2500\,\AA} & \colhead{4700\,\AA}
}
\startdata
0420.0$-$5022 &0.11 & $0.34\pm0.03$ &$9.6\pm1.3$ &\phn$8.3\pm0.8$  &\phn$7.2\pm0.7$&$4.20\pm0.22$& 1.2& 0/0   \\
0720.0$-$3125 &0.06 & $0.92\pm0.05$ &$4.6\pm0.3$ &\phn$6.1\pm0.4$  &\phn$8.6\pm0.5$&$3.43\pm0.12$& 1.4& 1.5/2\phn\phantom{.} \\
0806.4$-$4123 &0.10 & $0.22\pm0.02$ &$3.6\pm0.6$ &\phn$4.3\pm0.7$  &\phn$5.5\pm0.9$&$3.62\pm0.20$& 1.2& 0/0   \\
1308.6+2127   &0.10 & $0.19\pm0.01$ &$3.8\pm2.1$ &\phn$4.5\pm2.5$  &\phn$5.6\pm3.2$&$3.63\pm0.14$& 1.2& 0.4/1\phn\phantom{.} \\
1605.3+3249   &0.05 & $0.68\pm0.02$ &$5.0\pm0.5$ &\phn$7.2\pm0.8$  &   $11.6\pm1.3$&$3.23\pm0.07$& 1.3& 3.5/3\phn\phantom{.} \\
1856.5$-$3754 &0.05 & $4.25\pm0.07$ &$7.1\pm0.4$ &\phn$7.1\pm0.3$  &\phn$7.3\pm0.3$&$3.94\pm0.06$& 1.9& 2.6/2\phn\phantom{.} \\
2143.0+0654   &0.13 & $0.48\pm0.02$ &$9.5\pm0.4$ &$19.9\pm0.9$&        $49.8\pm2.3$&$2.53\pm0.08$    & 1.2& 0/0   \\
\enddata
\tablecomments{We give the nominal extinction $A_{V,\rm ref}$ (computed from $N_{\rm H}$
 in Table~\ref{tab:xray}), the absorbed flux density at 2500\,\AA, the power-law index
 $\alpha$ (defined such that $F_{\lambda}\propto \lambda^{-\alpha}$),
 and the $\chi^2$ and degrees of freedom for the fit.  We also list
 the excesses over the X-ray blackbody at 2500\,\AA\ (where the normalization is calculated)
 and at 1500\,\AA\ (appropriate for F140LP) and
 4700\,\AA\ (appropriate for F475W).  Uncertainties on the X-ray fits are
 included in the given uncertainties on the excess. Extinction at
 2500\,\AA\ is taken as $A_{2500\,{\rm \AA}}/A_V=2.32$.  We also
 give the change in power-law index with changes in extinction,
 $d\alpha/dA_V$, i.e., $\alpha(A_V)=\alpha(A_{V,{\rm
     ref}})+d\alpha/dA_V(A_V-A_{V,{\rm ref}})$.
}
\end{deluxetable*}

There exists an uncertainty in the models of strongly-magnetized NS
atmospheres at long wavelengths.  The plasma frequencies for the model
atmospheres constructed in, e.g., \citet{hpc08}, lie near the
optical/UV regime.  Classically, emission below the plasma frequency
should be suppressed, or at least highly modified.  The suppression
would be stronger at longer wavelengths (opposite to the behavior seen
in Fig.~\ref{fig:sed}), since the plasma frequency increases with
density.  Calculations using an ad-hoc approach to account for this
dense plasma suppression (see \citealt{hlpc03} for details) indicate
that deviations from Rayleigh-Jeans occur at far-UV/soft X-rays (for
the likely INS magnetic fields $B\gg 10^{12}$~G), while at optical
wavelengths, the flux is suppressed but the wavelength-dependence is
not affected.  Thus the above considerations may still apply.
Improvements on the treatment used here are possible (see, e.g.,
\citealt{brink80}; \citealt*{tzd04,pamp05};
\citealt{valpa05,hkc+07}) but do not qualitatively alter the results,
while a fully self-consistent model of the emission properties of the
condensed surface below the atmosphere is beyond the scope of this
work.

\ \\

Regardless of explanation, we have conclusively identified optical
counterparts to all 7 INSs and shown that the relatively
straightforward behavior shown by \rxjw\ is not necessarily the
dominant behavior.  Future modeling of these sources will have to
account for this diversity.  The quality of the \textit{HST} data has
allowed us to measure accurate SEDs and set references for future
astrometry.  

\acknowledgements We thank C.~Thompson, M.~Lyutikov, G.~Pavlov and the
STScI help desk for helpful discussions, and an anonymous referee for
useful comments.  Support for this work was provided by NASA (HST
award GO-11564.05).  Partial support was provided by the NSF under
grants PHY 05-51164 and AST 07-07633, by NASA under grant NNX08AX39G,
and by NSERC under a discovery grant to MHvK.  WCGH appreciates the
use of the computer facilities at the Kavli Institute for Particle
Astrophysics and Cosmology.  WCGH acknowledges support from the
Science and Technology Facilities Council (STFC) in the United
Kingdom.

{\it Facilities:} \facility{HST (ACS)}


\end{document}